\documentclass[sigconf, nonacm, screen]{acmart}

\usepackage[ruled,vlined]{algorithm2e}
\usepackage{algpseudocode}
\usepackage{multirow}
\usepackage{stackengine}
\usepackage{makecell}
\usepackage{tcolorbox}
\usepackage{colortbl}
\usepackage{diagbox}
\usepackage{enumitem}
\usepackage{bbding}
\usepackage{enumitem}

\newcommand{\squishlist}{
	\begin{list}{$\bullet$}
		{ \setlength{\itemsep}{1pt}
			\setlength{\parsep}{1pt}
			\setlength{\topsep}{2.5pt}
			\setlength{\partopsep}{0.5pt}
			\setlength{\leftmargin}{1em}
			\setlength{\labelwidth}{1em}
			\setlength{\labelsep}{0.6em}
		}
	}
	\newcommand{\squishend}{
	\end{list}
}

\newcommand{\rating}[1]{%
  \ifcase#1
    \FiveStarOpen\FiveStarOpen\FiveStarOpen\FiveStarOpen\FiveStarOpen % 0 stars
  \or
    \FiveStar\FiveStarOpen\FiveStarOpen\FiveStarOpen\FiveStarOpen % 1 star
  \or
    \FiveStar\FiveStar\FiveStarOpen\FiveStarOpen\FiveStarOpen % 2 stars
  \or
    \FiveStar\FiveStar\FiveStar\FiveStarOpen\FiveStarOpen % 3 stars
  \or
    \FiveStar\FiveStar\FiveStar\FiveStar\FiveStarOpen % 4 stars
  \or
    \FiveStar\FiveStar\FiveStar\FiveStar\FiveStar % 5 stars
  \fi
}

\usepackage{tikz}

\makeatletter
  \newcommand\figcaption{\def\@captype{figure}\caption}
  \newcommand\tabcaption{\def\@captype{table}\caption}
\makeatother

\makeatletter
\patchcmd{\@algocf@start}% <cmd>
  {-1.5em}% <search>
  {0pt}% <replace>
  {}{}% <success><failure>
\makeatother

\makeatletter
\def\blfootnote{\gdef\@thefnmark{}\@footnotetext}
\makeatother

\begin{document}

% \title{Is Hybrid Search All You Need?}
% \title{How to Develop and Optimize Your Hybrid Search System?}
\title{RTP-LLM: High-Performance Alibaba LLM Inference Engine}

%%
%% The "author" command and its associated commands are used to define the authors and their affiliations.
% \settopmatter{authorsperrow=4}

\author{$^{\dagger}$Boyu Tan, $^{\S}$Jiarui Guo, $^{\S}$Zongwei Lv, $^{\ddagger}$Haobo Sun, $^{\S}$Tong Yang, $^{\dagger}$Kan Liu, $^{\dagger}$Xinfei Shi, $^{\dagger}$Zetao Hu, $^{\dagger}$Yaxin Yu, $^{\dagger}$Chi Zhang, $^{\dagger}$Jianning Zhang, $^{\dagger}$Xi Yang, 
$^{\dagger}$Wei Zhang,
$^{\dagger}$Bo Cai,
$^{\dagger}$Silu Zhou,
$^{\dagger}$Xiyu Wang,
$^{\dagger}$Na He,
$^{\dagger}$Yinghao Yu,
$^{\dagger}$Wending Bao,
$^{\dagger}$Guiyang Huang,
$^{\dagger}$Yuxing Yuan,
$^{\dagger}$Juncheng Yin,
$^{\dagger}$Nan Wang,
$^{\dagger}$Lin Yang,
$^{\dagger}$Zechao Zhang,
$^{\ddagger}$Lu Chen,
$^{\dagger}$Guoding Li,
$^{\dagger}$Tao Lan,
$^{\dagger}$Lin Qu}
\affiliation{%
  \institution{
  {\large}$^{\dagger}$Alibaba Group $^{\S}$Peking University $^{\ddagger}$Zhejiang University\\
  {\large$^{\dagger}$\{tanboyu.tby, xinfei.sxf, huzetao.hzt, xieshui.yyx, yujing.zc, zhangjianning.zjn, wangyin.yx, zw193905, qisa.cb, silu.zsl, xiyu.wxy, yinghao.yyh, baowending.bwd, cangfei.hgy, yuanyuxing.yyx, yinjuncheng.yjc, kenan.wn,ziyu.yl, zhangzechao.zzc, guoding.lgd, tao.lant, xide.ql\}}@alibaba-inc.com ~~~$^{\dagger}$ luoli.hn@taobao.com $^{\S}$ \{ntguojiarui, lvzongwei, yangtong\}@pku.edu.cn
~~~$^{\ddagger}$\{haobosun, luchen\}@zju.edu.cn}
 \country{}
 }

%\author{Anonymous Author(s)}
\renewcommand{\shortauthors}{Boyu Tan et al.}
\begin{abstract}
Large Language Models (LLMs) have revolutionized AI applications, but deploying them at scale presents significant challenges. We present RTP-LLM, a high-performance inference engine for industrial-scale LLM deployment, successfully deployed across Alibaba Group serving over 100 million users. RTP-LLM addresses fundamental bottlenecks through integrated design. It optimizes model loading via file-order-driven I/O and parallel I/O-communication overlapping. The Prefill-Decode Disaggregation architecture decouples compute-intensive prefill from memory-bound decode phases, combined with hierarchical multi-tiered KV cache management enabling efficient cache reuse. In addition, RTP-LLM incorporates modular speculative decoding supporting multiple algorithms, adaptive KV cache quantization, and decoupled multimodal processing, with support for multi-level parallelism.

% Our comprehensive evaluations across diverse model architectures (8B-235B parameters) demonstrate RTP-LLM's superior performance compared with vLLM and SGLang: 4.7x-6.3x model loading speedup, 35-37\% TTFT P95 latency reduction with 215\% cache reuse improvement in production traffic scheduling, 1.12x-2.48x throughput improvement in speculative decoding, 1.86x-2.52x throughput improvement in multimodal inference, and 35-40\% batch latency reduction with 1.9x-3.0x TTFT improvement in quantized inference. RTP-LLM's production-proven architecture and open-source availability make it a comprehensive solution for industrial LLM deployment.

{Comprehensive evaluations across diverse model architectures (8B-235B parameters) have been conducted, where both controlled benchmarks and real production workloads are used.
The results demonstrate RTP-LLM's superior performance against vLLM and SGLang: 4.7x-6.3x model loading speedup, 35-37\% TTFT P95 latency reduction with 215\% cache reuse improvement in production traffic scheduling, 1.12x-2.48x and 1.86x-2.52x throughput improvements in speculative decoding and multimodal inference, respectively, and 35-40\% batch latency reduction with 1.9x-3.0x TTFT improvement in quantized inference. RTP-LLM's production-proven architecture and open-source availability make it a comprehensive solution for industrial LLM deployment.}

%\textbf{Keywords:} %Large Language Models, 
%Inference Serving, KV Cache Management %Prefill-Decode Disaggregation, KV Cache Management, Speculative Decoding, Distributed Inference, Production Systems

\end{abstract}

\maketitle
\blfootnote{Lu Chen is the corresponding author.}

% 原始结构，基本不改
\section{Introduction}
\label{sec:introduction}

The rapid advancement of Large Language Models (LLMs) has catalyzed a paradigm shift across industries, transforming applications ranging from conversational AI and code generation to enterprise automation \cite{gpt4, deepseekr1, llama3}. Modern models have reached a scale of hundreds of billions of parameters \cite{deepseekv3, qwen}, delivering unprecedented capabilities but also imposing severe computational and memory demands that traditional inference systems were never designed to handle. This chasm between model scale and deployability has emerged as a critical bottleneck, requiring fundamental rethinking of system architecture rather than incremental optimization.

The challenge is fundamentally rooted in the autoregressive nature of LLM inference, where each generated token depends on all preceding tokens through sequential attention computations \cite{vaswani2017attention}. Unlike traditional machine learning workloads that benefit from embarrassingly parallel batch processing, autoregressive generation creates an intrinsic sequential bottleneck that fundamentally limits GPU utilization and throughput \cite{yang2024queueing, agrawal2024taming}. Compounding this issue, the Key-Value (KV) cache—a critical data structure for storing intermediate attention states—grows dynamically during generation and can dominate memory consumption, particularly as context lengths exceed 128K tokens \cite{zhang2023h2o, zhao2025memo, deng2025alayadb}. Early inference frameworks treat these constraints as immutable, resulting in suboptimal resource utilization and prohibitive latency for production deployments.

Recent years have witnessed significant innovations targeting specific facets of this problem. Systems like \textbf{vLLM} introduced PagedAttention, revolutionizing memory management by treating the KV cache as a paged virtual memory system \cite{vllm}. \textbf{TensorRT-LLM} delivered kernel-level optimizations tailored to NVIDIA hardware \cite{tensorrtllm}, while \textbf{FlashAttention} restructured attention computations to reduce memory bandwidth pressure \cite{dao2022flashattention, dao2023flashattention}. Despite these advances, a critical gap persists: most existing solutions optimize isolated components while neglecting the systemic interactions required for production-scale deployment. They typically focus on single-node performance, lack support for heterogeneous hardware, and fail to address the full spectrum of challenges—from dynamic workload scheduling to rapid model iteration—that enterprises face in practice \cite{kamath2024llms, li2024llm, jo2025sparellm}.

Production LLM deployments face four fundamental challenges that existing systems inadequately address:

\textbf{Challenge I: Underutilized GPUs under dynamic sequential workloads.} Request patterns exhibit extreme variability in input length (from short queries to 128K+ contexts), output length, and computational intensity \cite{wang2025burstgpt, agrawal2025evaluating}. The memory-bound nature of autoregressive attention leaves compute units idle, while static batching strategies cannot adapt to request dynamism, resulting in unpredictable latency and suboptimal throughput.

\textbf{Challenge II: Memory exhaustion from unconstrained KV cache growth.} The KV cache grows linearly with sequence length and batch size, quickly becoming the dominant memory consumer \cite{liu2025clusterkv, lee2024infinigen}. Traditional memory allocators struggle with fragmentation and inefficient sharing across requests with diverse context lengths. As modern models support ever-longer contexts, memory management has become a hard capacity constraint that limits concurrency and scalability.

\textbf{Challenge III: System rigidity in the face of architectural heterogeneity.} Modern model architectures introduce significant complexity that existing systems struggle to handle efficiently. Large-scale Mixture-of-Experts (MoE) models exceeding 600B parameters require efficient expert routing and rapid weight loading \cite{deepseekv3}, while multi-modal models combine vision encoders and language models that have fundamentally different computational characteristics and execution patterns, requiring coordinated execution across heterogeneous computation components.

\textbf{Challenge IV: Operational fragility impeding rapid iteration.} Enterprise deployments demand minute-level loading of 600B+ parameter models to enable continuous updates across business units \cite{pothukuchi2025llmops}. Existing systems require hours for weight loading and lack production-grade fault tolerance, rolling update mechanisms, and per-request performance isolation needed to meet stringent latency SLOs under fluctuating load.

To address these challenges, we present \textbf{RTP-LLM}, a holistic inference system developed by Alibaba's Foundation Model Inference Team and battle-tested across production deployments serving Taobao, Tmall, and Cainiao. %RTP-LLM integrates multiple optimization techniques into a cohesive, production-ready platform.

To enable rapid model iteration (addressing Challenge IV), RTP-LLM implements \textbf{Optimized Model Loading} (Section~\ref{sec: model_load}) through file-order-driven I/O, shared memory reuse, and parallel I/O-commu-nication overlap, achieving 1.4x-6.3x faster loading times compared to vLLM and SGLang, 
enabling minute-level deployment of 600B+ parameter models. RTP-LLM incorporates enterprise-grade operational features including fault tolerance, rolling updates, and per-request performance isolation to ensure predictable latency under stringent SLOs.

To maximize GPU utilization and memory efficiency (addressing Challenge I and II), RTP-LLM implements \textbf{Prefill-Decode Disaggregation} \cite{qin2025mooncake, patel2024splitwise, zhong2024distserve} (Section~\ref{sec:traffic}) that physically decouples compute-intensive prefill from memory-bound decode phases, enabling independent scaling and optimal resource allocation. \textbf{Dynamic Traffic Scheduling} with intelligent load balancing continuously reprioritizes requests based on queue state, KV cache footprint, and latency targets, maximizing GPU utilization through dynamic batching \cite{yu2022orca, he2024deferred}. \textbf{Hierarchical Multi-Tiered KV Cache Management} (Section~\ref{sec:traffic}) spans GPU memory, local CPU memory, remote CPU memory via RDMA, and distributed storage, with unified hash-based prefix matching enabling efficient cache reuse. \textbf{Prefix Caching} enables fine-grained reuse of KV cache pages across requests sharing common prefixes (e.g., system prompts, RAG passages), significantly reducing computational overhead \cite{zheng2024batchllm, jin2024ragcache, sglang}. Production evaluations demonstrate 35-37\% TTFT P95 latency reduction and 215\% cache reuse improvement, enabling 75\% reduction in prefill machine count.

To break the sequential generation bottleneck (further addressing Challenge I), we employ \textbf{Multi-Token Speculative Decoding} (Section~\ref{sec:speculative_decoding}) supporting multiple algorithms (Medusa, Eagle, Prompt Lookup) that achieve 1.12x-2.48x throughput improvement by predicting and verifying multiple future tokens in parallel while preserving output quality \cite{spector2023accelerating, xia2024unlocking}.

To support diverse model architectures (addressing Challenges II and III), RTP-LLM implements \textbf{Adaptive KV Cache Quantization} (Section~\ref{sec:extension}) that reduces memory footprint and improves the performance. \textbf{Multi-Level Parallelism} (Section~\ref{sec:extension}) integrates Tensor, Data, Pipeline, and Expert Parallelism to support diverse model architectures from dense models to MoE models exceeding 600B parameters. For multimodal models, \textbf{Decoupled ViT-LLM Processing} (Section~\ref{sec:extension}) separates vision encoding from language generation, achieving 1.86x-2.52x throughput improvement and 2.12x-2.36x TTFT reduction.

% To enable rapid model iteration (addressing Challenge IV), RTP-LLM implements \textbf{Optimized Model Loading} (Section~\ref{sec: model_load}) through file-order-driven I/O, shared memory reuse, and parallel I/O-communication overlap, achieving 1.4x-6.3x faster loading times compared to vLLM and SGLang, enabling minute-level deployment of 600B+ parameter models. RTP-LLM incorporates enterprise-grade operational features including fault tolerance, rolling updates, and per-request performance isolation to ensure predictable latency under stringent SLOs.

This paper makes the following contributions:
\begin{itemize}[topsep=0pt, leftmargin=*]
    \item We present the design and implementation of RTP-LLM, a comprehensive inference system that integrates memory management, scheduling, and hardware acceleration into a cohesive, production-ready platform. RTP-LLM addresses the full inference stack with battle-tested reliability across Alibaba's ecosystem, serving over 100 million users in production environments.

    \item We describe system design and engineering practices including hierarchical load balancing for disaggregated serving, unified multi-modal request orchestration, and adaptive resource allocation schemes that dynamically adjust GPU memory fractions between prefill and decode engines based on live traffic analysis.

    \item We provide extensive performance evaluations across diverse model architectures (dense, MoE, multi-modal) using both controlled benchmarks and \textbf{real production workloads}, offering actionable insights into the effectiveness and interaction of different optimization techniques in real-world deployment.

    % \item We share lessons learned from large-scale production deployments across hundreds of thousands of GPUs, codifying best practices for model deployment, capacity planning, and performance tuning that transcend specific hardware or model choices.

    \item We release RTP-LLM as open-source software \cite{source}, providing an extensible foundation for research and development in LLM inference. The system has already attracted active community adoption, fostering further innovation in the field.
\end{itemize}

%Our evaluation (Section~\ref{sec: eval setup}) use both standard benchmarks and {real production deployments}: traffic scheduling and PD disaggregation use production traffic and configurations from our live services; speculative decoding and MoE evaluation use data collected from actual deployment scenarios. 
%The rest of this paper is organized as follows. Section~\ref{sec: framework} presents RTP-LLM's overall system architecture and core components. Section~\ref{sec: model_load} details optimized model loading techniques. Section~\ref{sec:traffic} describes Prefill-Decode Disaggregation, dynamic traffic scheduling, and hierarchical KV cache management. Section~\ref{sec:speculative_decoding} presents the speculative decoding framework. Section~\ref{sec:extension} details system extensions including multi-level parallelism, quantization techniques, and multimodal model support. Section~\ref{sec: eval setup} presents comprehensive performance benchmarks and production metrics. Finally, Section~\ref{sec:conclusion} concludes this paper with future work directions.

\section{Background}
\label{sec: background}
%The deployment of LLMs in production environments has emerged as one of the most critical challenges in modern artificial intelligence systems \cite{pothukuchi2025llmops}. 
As model sizes (i.e., \#parameters) of LLMs have grown from millions \cite{brown2020language, radford2019language, koroteev2021bert} to hundreds of billions  \cite{qwen, deepseekv3}, the computational and memory requirements for real-time inference have escalated exponentially, making it challenging to deploy LLMs in production environments.
%necessitating fundamental innovations in system architecture and optimization techniques.
The current landscape of LLM inference systems is characterized by a diverse ecosystem of solutions, each addressing different aspects of the inference challenge \cite{tgi, wang2023tabi}. %Traditional model serving frameworks, originally designed for conventional machine learning workloads, have proven inadequate for the unique characteristics of autoregressive text generation. This inadequacy has driven the development of specialized inference engines that can handle the sequential nature of token generation while maintaining high throughput and low latency.

The fundamental challenge in LLM inference stems from the autoregressive nature of text generation, where each token depends on all previously generated tokens \cite{vaswani2017attention}. This sequential dependency limits the parallelization used in traditional batch processing systems \cite{you2024linear, hilsenbek2024breaking}. Early inference systems treated this sequential constraint as immutable, leading to suboptimal resource utilization and poor scalability characteristics.

% The breakthrough came with the recognition that while token generation must remain sequential, the underlying computations could be parallelized through sophisticated attention mechanisms and memory management techniques \cite{dao2022flashattention, dao2023flashattention}. This insight led to the development of techniques such as PagedAttention \cite{vllm}, which enables efficient memory management for variable-length sequences, and continuous batching \cite{yu2022orca}, which allows dynamic request scheduling without compromising generation quality.

% Memory management represents one of the most critical challenges in LLM inference systems \cite{hatalis2023memory, zhang2025jenga}. The Key-Value (KV) cache, essential for efficient attention computation, grows dynamically during generation and can consume substantial memory resources \cite{liu2025clusterkv, lee2024infinigen}. Traditional memory allocation strategies, designed for fixed-size allocations, prove inadequate for the dynamic and variable-size memory requirements of LLM inference. The development of paged memory management systems has revolutionized memory efficiency in LLM inference, treating KV cache as a collection of fixed-size pages that can be allocated, deallocated, and shared across different requests.

The breakthrough came with the recognition that while token generation must remain sequential, the underlying computations
could be parallelized through sophisticated attention mechanisms and memory management techniques \cite{dao2022flashattention, dao2023flashattention}. This insight led to the development of techniques such as continuous batching \cite{yu2022orca} which allows dynamic request scheduling without compromising generation quality. However, memory management represents one of the most critical challenges in LLM inference systems \cite{hatalis2023memory, zhang2025jenga, gao2025apt}. The Key-Value (KV) cache, essential for efficient attention computation, grows dynamically during generation and can consume substantial memory resources \cite{liu2025clusterkv, lee2024infinigen}. Traditional memory allocation strategies, designed for fixed-size allocations, prove inadequate for the dynamic and variable-size memory requirements of LLM inference. To address these challenges, paged memory management systems, such as PagedAttention \cite{vllm}, have emerged as a revolutionary approach, treating KV cache as a collection of fixed-size pages that can be allocated, deallocated, and shared across different requests, thereby enabling efficient memory management for variable-length sequences and significantly improving memory efficiency and utilization.

Major cloud providers have developed proprietary inference platforms optimized for their specific infrastructure and hardware configurations \cite{vllm, sglang, tensorrtllm}. These platforms often achieve impressive performance metrics through extensive optimization and specialized hardware, but suffer from vendor lock-in and limited customization capabilities. The open source ecosystem has produced several notable inference engines, with vLLM \cite{vllm} representing a significant advancement in the field. However, vLLM and similar systems face limitations in production environments, particularly in their focus on single-node optimization and limited support for heterogeneous hardware environments.

\begin{figure}
  \centering
  \includegraphics[width=\linewidth]{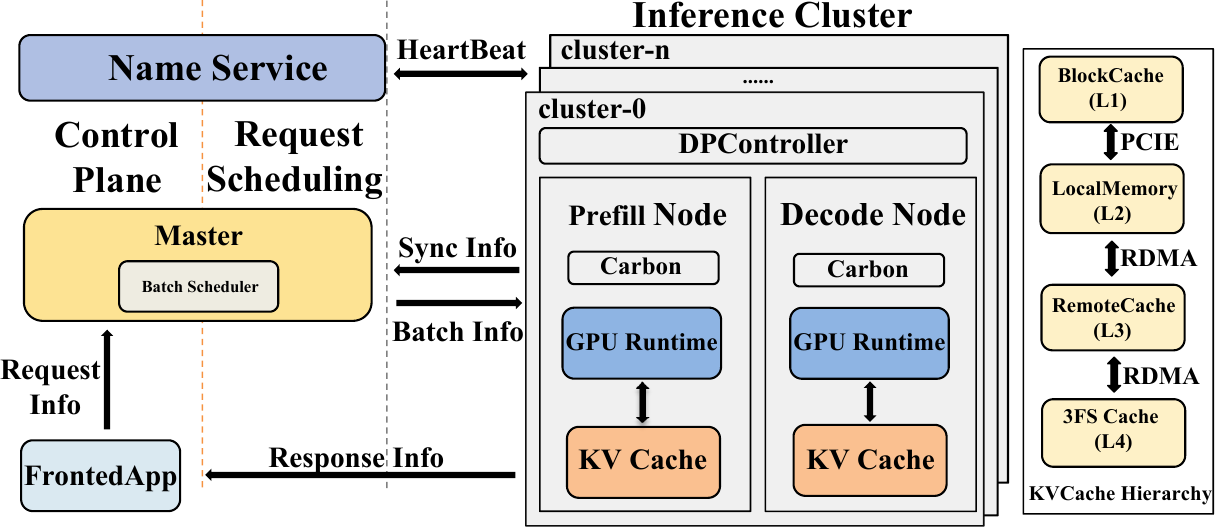}
  \vspace{-6mm}
  \caption{RTP-LLM System Architecture}
  \label{fig:rtp_llm_framework}
  \vspace{-6mm}
\end{figure}

{The LLM inference process can be decomposed into two distinct phases with different computational characteristics. The \textbf{Prefill Phase} processes the entire input prompt in parallel, generating and storing Key-Value (KV) Cache entries for all prompt tokens, and produces the first output token. This phase involves parallel computation across all input tokens, making it compute-bound. The \textbf{Decode Phase} generates subsequent tokens autoregressively, utilizing the current token and historical KV Cache. Each decode iteration processes a single token and updates the KV Cache accordingly, making it memory bandwidth-bound. \textbf{Prefill-Decode Disaggregation} exploits this computational asymmetry by physically decoupling these two phases onto dedicated computational resources, enabling independent scaling: prefill engines maximize throughput via large batches while decode engines optimize for low-latency memory access \cite{patel2024splitwise, qin2025mooncake, zhong2024distserve}.
}

{\textbf{Speculative Decoding} addresses the sequential decode bottleneck by introducing parallel token verification mechanisms \cite{spector2023accelerating}. The approach employs a dual-model architecture: a \textbf{Propose Model ($M_q$)} that generates $k$ candidate tokens, and a \textbf{Score Model ($M_p$)} that validates proposed tokens through parallel scoring. The process operates through three stages: (1) Propose stage generates $k$ candidates, (2) Score stage evaluates all $k$ tokens in parallel, and (3) Verification stage determines acceptance based on probability distributions. This transforms sequential decode into parallel verification, significantly improving GPU utilization. For single-request scenarios with negligible verification overhead, when using a rule-based propose model, the theoretical speedup is proportional to the ratio of scoring $k$ tokens versus sequential decoding, with empirical results demonstrating substantial speedup in ideal conditions. As request concurrency increases, speculative sampling's latency benefits diminish due to computational overhead from rejected tokens and resource contention. However, it remains beneficial in memory-constrained environments, scenarios with framework limitations preventing maximum concurrency, and long sequence optimizations requiring improved KV Cache access patterns.
}

Production LLM deployments require high scalability.
%challenges that differ significantly from traditional machine learning serving scenarios. 
The dynamic nature of LLM workloads, characterized by high variability in input length, output length, and computational requirements, resulting in complex scheduling and resource allocation. Modern production environments often involve diverse hardware configurations, from high-end NVIDIA GPUs to AMD ROCm systems and custom accelerators, each requiring specialized implementations for optimal performance.
The complexity of LLM inference optimization requires comprehensive solutions that integrate multiple optimization techniques into cohesive systems. % Isolated optimizations, while potentially effective in specific scenarios, often fail to provide the consistent performance improvements required in production environments. % The development of comprehensive solutions requires extensive experience with production deployments, deep understanding of the interactions between different optimization techniques, and sophisticated system design capabilities.

RTP-LLM represents such a comprehensive solution, developed through extensive production deployment experience across Alibaba's ecosystem and designed to address the multifaceted challenges of LLM inference optimization. %The system's holistic approach to optimization, combined with its production-proven reliability and open source nature, positions it as a significant contribution to both the research community and production practitioners seeking robust, scalable LLM inference solutions.

 % \vspace{-6mm}
\section{RTP-LLM System Design}
\label{sec: framework}
In this section, we present a high-level overview of the comprehensive architectural design of the RTP-LLM inference framework. 
% \vspace{-7mm}
\subsection{Core System Components}

Figure \ref{fig:rtp_llm_framework} illustrates the main components of RTP-LLM, which include the Frontend Application, the Master, the Prefill Node, the Decode Node, a Multi-Tiered Cache, the Name Service, and the DP-Controller.

The Frontend Application serves as the entry point, tasked with accepting user requests and subsequently returning the corresponding responses. It handles request preprocessing, including tokenization and metadata extraction, before forwarding requests to the Master component for scheduling and execution. The Name-Service performs heartbeat detection and service discovery to ascertain which deployed clusters are operational. It is not responsible for load balancing. Instead, the Master component assumes the critical role of traffic scheduling and global coordination. The Master maintains a global view of system state, including worker availability, KV Cache distribution, and current load conditions, enabling optimal scheduling decisions that maximize throughput while meeting latency requirements.

Regarding practical deployment, our framework supports two distinct strategies: PD-Fusion and PD-Disaggregation. The former, PD-Fusion, refers to a deployment mode where both the prefill and decode phases of the inference process are co-located and executed within a single, unified node. In contrast, the latter, PD-Disaggregation—which is the topology depicted in Figure \ref{fig:rtp_llm_framework}—physically decouples these two computational stages into dedicated nodes. Each operational inference service node is accompanied by a dedicated Carbon service, which is responsible for the automatic recovery and restart of the inference service in the event of a failure.

The Multi-Tiered Cache is a specialized component architected to enhance cache utilization efficiency throughout the inference pipeline. It implements a hierarchical storage system spanning GPU memory, local CPU memory, remote CPU memory via RDMA, and distributed storage, enabling efficient KV cache reuse across requests while minimizing memory bandwidth pressure.

% Finally, the DP-Controller is responsible for orchestrating and managing the execution of a batch of requests within a singular deployment context.

Finally, the DP-Controller is responsible for orchestrating and managing the execution of a batch of requests within a singular deployment context. It coordinates with the Master to receive scheduled batches and manages local resource allocation, including GPU memory management and batch execution. The Prefill Node and Decode Node represent the core inference execution components. Prefill Nodes handle the compute-intensive prefill phase, processing entire input prompts in parallel and generating initial KV cache states. Decode Nodes manage the memory-bound decode phase, generating tokens autoregressively using the cached attention states. In PD-Disaggregation deployments, these nodes operate independently, allowing for specialized optimization and independent scaling of each phase.

\setlength{\textfloatsep}{4pt}
\begin{algorithm} [tb]
\caption{\textsc{RTP-LLM Hierarchical Architecture}}
\label{alg:rtp_llm_hierarchical_workflow}
\LinesNumbered
\KwIn{User request $req$, Cluster state $cluster\_state$}
\KwOut{Processing result $result$}

\tcp{Generate hash keys for prefix matching}
$\mathcal{H} \gets$ GenerateHashKeys($req.tokens$)
% \textcolor{blue}{\Comment{\textsf{Generate hash keys for prefix matching}}}

\tcp{Find matched blocks in cache}
$\mathcal{M} \gets$ PrefixMatching($\mathcal{H}$)
% \textcolor{blue}{\Comment{\textsf{Find matched blocks in cache}}}

% \If{$\mathcal{M} \neq \emptyset$}{
%     $req.kv\_cache \gets$ LoadFromCache($\mathcal{M}$) \textcolor{blue}{\Comment{\textsf{Cache hit, reuse existing blocks}}}
% }
% \Else{
%     $req.kv\_cache \gets$ AllocateNewCache($req.length$) \textcolor{blue}{\Comment{\textsf{Cache miss, allocate new blocks}}}
% }

\tcp{Master load balancing and batching}
$\mathcal{B} \gets$ MasterDecision($req$, $cluster\_state$)
% \textcolor{blue}{\Comment{\textsf{Master load balancing and batching}}}

\ForAll{$req \in \mathcal{B}$}{
    \If{$req.kv\_cache \in block\_cache$}{
        \tcp{BlockCache layer: GPU memory}
        UpdateReferenceCount($req.kv\_cache$)
        % \textcolor{blue}{\Comment{\textsf{BlockCache layer: GPU memory}}}
    }
    \ElseIf{$req.kv\_cache \in local\_memory$}{
        \tcp{LocalMemory layer: Local CPU memory}
        LoadToGPU($req.kv\_cache$)
        % \textcolor{blue}{\Comment{\textsf{LocalMemory layer: Local CPU memory}}}
    }
    \ElseIf{$req.kv\_cache \in remote\_cache$}{
        \tcp{RemoteMemory layer: Remote CPU memory}
        RDMATransfer($req.kv\_cache$, $local\_memory$)
        % \textcolor{blue}{\Comment{\textsf{RemoteMemory layer: Remote CPU memory}}}
    }
    \Else{
        \tcp{Remote3fs layer: Distributed storage}
        LoadFrom3FS($req.kv\_cache$, $remote\_cache$)
        % \textcolor{blue}{\Comment{\textsf{Remote3fs layer: Distributed storage}}}
    }
}

\tcp{Execute inference on batch}
$result \gets$ ExecuteInference($\mathcal{B}$)
% \textcolor{blue}{\Comment{\textsf{Execute inference on batch}}}

\tcp{Return cache and update LRU}
CacheReturnAndUpdate($\mathcal{B}.kv\_cache$)
% \textcolor{blue}{\Comment{\textsf{Return cache and update LRU}}}

\textbf{return} $result$

\end{algorithm}
% \vspace{-4mm}

\vspace{-2mm}
\subsection{Execution Stream}

The complete system workflow of RTP-LLM is outlined in the pseudocode provided in Algorithm \ref{alg:rtp_llm_hierarchical_workflow}. Initial user requests are directed to one of the available FrontedApp instances (which operates in a load-balanced configuration). Each FrontedApp performs initial request preprocessing and metadata extraction before synchronously forwarding the comprehensive request payload to the centralized Master node. \textbf{The Master Node initiates the request processing workflow} by generating the requisite \textbf{prefix hash keys ($\mathcal{H}$)} from the incoming user request (Algorithm 1, Line 1: \texttt{GenerateHashKeys}). The Master node then utilizes these generated hash keys ($\mathcal{H}$) to perform \textbf{prefix matching} against the global cache, identifying the set of matched candidate KV cache blocks ($\mathcal{M}$) (Algorithm 1, Line 2: \texttt{PrefixMatching}). Next, the Master node performs decisive \textbf{load analysis and scheduling} based on the request details, the matching results ($\mathcal{M}$), and the current $\texttt{cluster\_state}$, formulating an execution batch ($\mathcal{B}$) (Algorithm 1, Line 3: \texttt{MasterDecision}). This batch is subsequently dispatched to the designated Inference Node for execution. 

The core of the system's efficiency is implemented through a \textbf{four-tier hierarchical memory access mechanism} (Algorithm 1, Lines 4--12). For every request $req \in \mathcal{B}$, the system attempts to load the necessary KV cache blocks (\texttt{req.kv\_cache}) from the fastest to the slowest storage tiers:
\begin{enumerate}[topsep=2pt, leftmargin=*]
    \item \textbf{GPU Memory ($\texttt{block\_cache}$):} If the block is present, its reference count is updated (\texttt{UpdateReferenceCount}).
    \item \textbf{Local CPU Memory ($\texttt{local\_memory}$):} If absent from the GPU, the block is loaded from the local memory to the GPU (\texttt{LoadToGPU}).
    \item \textbf{Remote CPU Memory ($\texttt{remote\_cache}$):} If absent locally, the block is transferred from remote CPU memory to local CPU memory using high-speed \textbf{RDMA} (\texttt{RDMATransfer}).
    \item \textbf{Distributed Storage ($\texttt{Remote3fs}$):} If the block is not found in any higher-tier memory, it is retrieved from the centralized distributed storage and placed into the remote cache (\texttt{LoadFrom3FS}).
\end{enumerate}

This hierarchical memory access mechanism ensures that all KV cache dependencies are fulfilled and staged onto the GPU before the inference step. After that, the Inference Node performs the \textbf{model execution} on the batch ($\mathcal{B}$) (Algorithm 1, Line 13: \texttt{ExecuteInference}). Upon completion, the utilized KV cache blocks are returned and their metadata updated via \texttt{CacheReturnAndUpdate} (Algorithm 1, Line 14), which specifically includes the critical step of updating the \textbf{Least Recently Used (LRU)} metrics for effective cache management across all memory tiers. The final result is then returned (Algorithm 1, Line 15). 

The Inference Node subsequently delivers the final result directly back to the originating FrontedApp to complete the request cycle.

% 以下内容重新组织结构，主要调换顺序，不大改内容

\begin{figure}
  \centering
  \includegraphics[width=\linewidth]{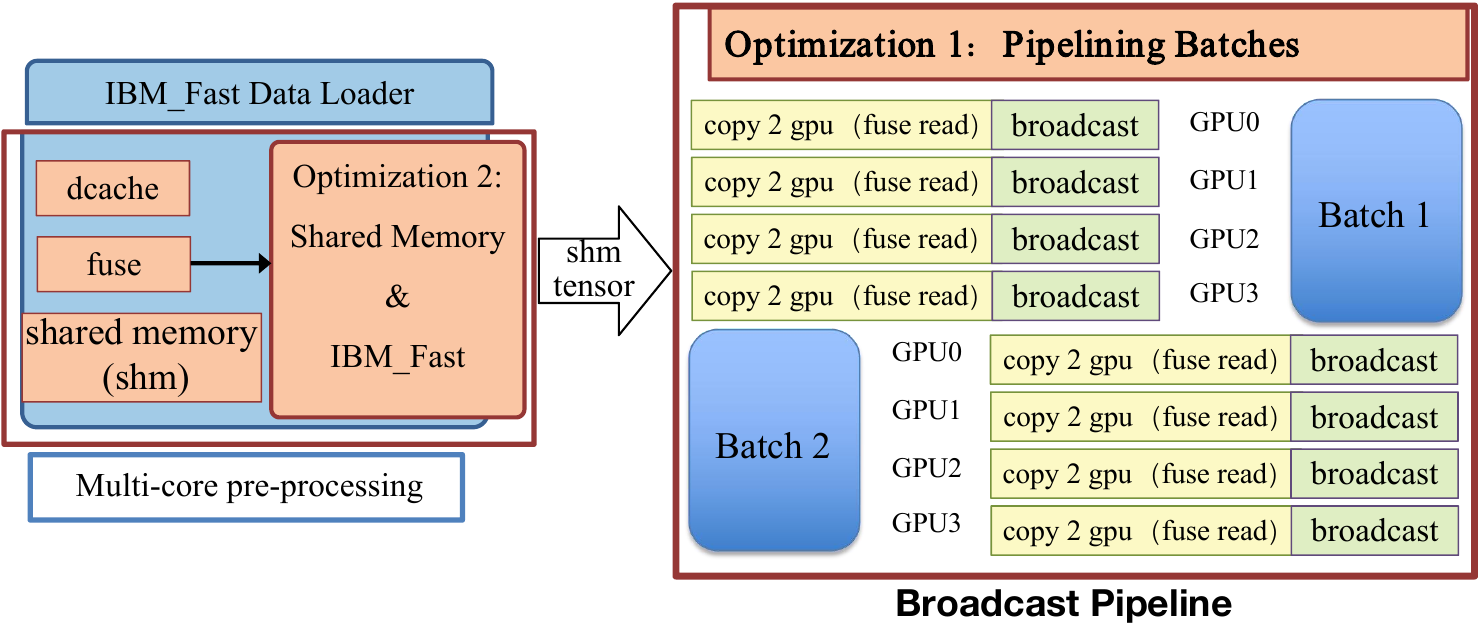}
  \vspace{-7mm}
  \caption{Model Load Optimizations}
  \label{fig:model_load}
  % \vspace{-3mm}
\end{figure}

\vspace{-2mm}
\section{Efficient Model Loading}

\label{sec: model_load}
LLM loading presents significant performance challenges in distributed environments, particularly when dealing with models exceeding hundreds of billions of parameters across multiple tensor parallel processes. Critically, in the Alibaba production environment, all model checkpoints are stored on an internal FUSE (Filesystem in Userspace) cloud storage system, which is mounted on the inference nodes. This infrastructure imposes a strict constraint: the efficiency of file access is heavily dependent on I/O patterns.

For traditional model-structure-driven loading approach, each tensor parallel process reads all model files to extract only its assigned tensor portions. This creates two critical performance problems: redundant file reads across processes and non-sequential file access patterns that severely degrade FUSE prefetching efficiency and file-level caching utilization.

RTP-LLM addresses the above performance problems through a comprehensive optimization strategy that tackles both I/O efficiency and memory management bottlenecks. Our strategy centers on restructuring the loading paradigm from model-structure-driven to file-order-driven loading. Instead of iterating through weight modules and loading their constituent tensors, we iterate through model files sequentially, loading all tensors from each file before proceeding to the next. This approach ensures sequential file access patterns that maximize FUSE prefetching effectiveness while maintaining compatibility with community engines.

 Figure \ref{fig:model_load} highlights our key optimizations. To eliminate redundant reads, we integrate IBM's fastsafetensors library \cite{yoshimura2025speedingmodelloadingfastsafetensors} with RTP fastsafetensors implementation. The hybrid approach assigns each model file to a single process for reading, then leverages PyTorch distributed broadcast to efficiently share tensors across all processes. This eliminates the need for each process to read every file while maintaining the benefits of FUSE-based I/O optimizations.

Memory management optimization reduces the significant overhead of repeated shared memory allocation. The original fastsafetensors library allocates and registers pinned memory for each file read operation, requiring 600ms overhead per 2GB allocation. We encapsulate the loading interface within a class that reuses a single shared memory buffer across multiple file reads, eliminating this redundant allocation cost.

Finally, we optimize communication and I/O overlap by parallelizing file reading operations with tensor broadcasting. Rather than sequentially reading files followed by tensor distribution, we overlap these two phases (i.e., parallelizing file reading and tensor distribution) in order to reduce overall loading latency. This is achieved by allowing file I/O operations to proceed concurrently with tensor broadcasting, maximizing resource utilization across the distributed system.

The combination of file-ordered loading, hybrid distributed reading, shared memory reuse, and parallel I/O-communication overlapping creates a comprehensive solution that improves the performance of large-scale model loading while maintaining compatibility with existing model formats and community frameworks.

\section{Traffic Scheduling}

\label{sec:traffic}

% \gjr{
% In this section, we outline the traffic scheduling mechanisms of RTP-LLM. 
% We first implement Prefill-Decode Disaggregation as the foundation of our scheduling framework. 
% Then we describe the Load Balance Strategy of RTP-LLM, which dynamically redistributes requests based on real-time workloads. 
% Finally we introduce KV cache Management to enable efficient memory utilization and consistent serving performance across workers. 
% }

{In this section, we outline the traffic scheduling mechanism.
We describe the Load Balance Strategy of RTP-LLM, which dynamically redistributes requests based on real-time workloads.
Next, we introduce KV cache Management to enable efficient memory utilization and consistent serving performance across workers.}

% \subsection{Preliminary: Prefill-Decode Disaggregation}

% The LLM inference process can be decomposed into two distinct phases:

% \begin{itemize}[leftmargin=1em]
%     \item \textbf{Prefill Phase}: During this stage, the model processes the entire input prompt in parallel, generating and storing Key-Value (KV) Cache entries for all prompt tokens, and produces the first output token. This phase involves parallel computation across all input tokens.

%     \item \textbf{Decode Phase}: Following the prefill phase, the model generates subsequent tokens autoregressively, utilizing the current token and historical KV Cache. Each decode iteration processes a single token and updates the KV Cache accordingly.
% \end{itemize}

% The computational characteristics of these phases differ significantly. The prefill phase typically processes hundreds of tokens simultaneously, making it compute-bound, while the decode phase processes individual tokens sequentially, making it memory bandwidth-bound.

\subsection{Load Balancing Strategy}

RTP-LLM employs distinct load balancing mechanisms for Prefill and Decode stages to optimize throughput and latency.

For Prefill requests, FrontApp tokenizes input sequences and generates block hash identifiers. Each request is divided into blocks (e.g., 64 tokens per block), and each block's hash key is computed based on its token IDs. FrontApp sends to Master the request's block hash IDs, sequence length, and optional chat ID.

% Master groups requests with similar sequence lengths into batches to minimize padding overhead. Window size $w = \max(DP\_size, |R|)$ is dynamically adjusted based on queue depth $|R|$ and DP group size. Master queries DP-Controllers for real-time load status, including running/waiting requests, GPU memory utilization, and KV cache occupancy.

{Master groups requests with similar sequence lengths into batches to minimize padding overhead. Window size $w = \max(DP\_size, |R|)$ is dynamically adjusted based on DP group size and queue depth. Master queries DP-Controllers for real-time load status, including running/waiting requests, GPU memory, and KV cache occupancy}.

When all DP-Controllers are busy, Master employs predictive scheduling by estimating completion times:
\begin{equation}
t_{available}(d_i) = \max_{r \in running(d_i)} t_{start}(r) + \hat{t}_{prefill}(r)
\end{equation}
% where $\hat{t}_{prefill}(r)$ is predicted based on sequence length and batch composition. Master schedules requests to the DP-Controller expected to complete first, reducing queue wait time.
{where $t_{start}(r)$ denotes the start time when request $r$ began execution on DP-Controller $d_i$, and $\hat{t}_{prefill}(r)$ is predicted based on sequence length and batch composition. Master schedules requests to the DP-Controller expected to complete first, in order to reduce the queue wait time.}

Decode requests prioritize KV cache affinity: when a request arrives with a chat ID, Master checks if this chat session was previously assigned to a worker. If a match exists and the worker has sufficient capacity, Master directly routes to that worker, exploiting local KV cache locality. For cache management, Master implements admission control, eviction priority, and backpressure signaling to prevent cache thrashing.

\subsection{KV Cache Management}

RTP-LLM employs a hash-based approach for efficient prefix matching across workers. The Local KV Cache Manager maintains a unified hash map that aggregates cache keys from all workers, mapping hash keys to block identifiers and worker metadata.

\subsubsection{Local KV Cache Manager}

The Local KV Cache Manager aggregates cache keys from all workers into a unified hash map structure. Each entry in the map stores:
\begin{itemize}[leftmargin=1em]
    \item \textbf{Hash key}: Block hash identifier computed from token IDs in the block
    \item \textbf{Block ID}: Block identifier on the worker
    \item \textbf{Worker cache info}: Set of $(worker\_id, cache\_metadata)$ pairs indicating which workers have cached this block
\end{itemize}

Instead of maintaining separate hash maps for each worker requiring $O(B \times W)$ lookups, we merge cache keys from all workers into a single hash map, enabling $O(B)$ complexity for prefix matching, where $B$ is the number of blocks and $W$ is the number of workers.

Master queries worker status at high frequency (20ms) for scheduling decisions, while cache key synchronization operates at lower frequency (50ms). Workers maintain cache version numbers. When requesting cache keys, the manager includes the last known version. If unchanged, workers return lightweight acknowledgment. If changed, workers return delta updates to minimize data transfer.

\subsubsection{Prefix Cache Matching}

When FrontApp sends a request with block hash IDs $H = [h_1, h_2, \ldots, h_B]$ (computed from token sequences), Master queries the Local KV Cache Manager for prefix matching. The matching process queries the unified hash map:

% \begin{enumerate}
%     \item Initialize matched length $l = 0$
%     \item For each block $i$ from 1 to $B$:
%     \begin{itemize}
%         \item Query the hash map with hash key $h_i$
%         \item If the hash key exists in the map, retrieve associated worker cache info
%         \item If match succeeds, increment $l$ and update match results for all workers cached at this hash key: $M[worker\_id] = \max(M[worker\_id], l)$
%         \item If hash key not found, terminate matching
%     \end{itemize}
%     \item Return mapping $M: WorkerID \rightarrow MatchLength$ indicating the maximum matching prefix length for each worker
% \end{enumerate}

\begin{algorithm}[t]
\caption{\textsc{Prefix Cache Matching}}
\label{alg:prefix_cache_matching}
\LinesNumbered
\KwIn{Block hash IDs $H = [h_1, h_2, \ldots, h_B]$, Unified hash map $\mathcal{H}$}
\KwOut{Mapping $M: WorkerID \rightarrow MatchLength$}

$l \gets 0$\;
$M \gets \emptyset$\;

\For{$i = 1$ \textbf{to} $B$}{
    \If{$h_i \in \mathcal{H}$}{
        $worker\_info \gets \mathcal{H}[h_i]$; \space \tcp{Get workers}
        % \textcolor{blue}{\Comment{Get workers}}\;
        $l \gets l + 1$\;
        \ForAll{$w \in worker\_info$}{
            $M[w] \gets \max(M[w], l)$; \space \tcp{Update max length}
            % \textcolor{blue}{\Comment{Update max length}}\;
        }
    }
    \Else{
        \textbf{break} \space \tcp{Terminate}
        % \textcolor{blue}{\Comment{Terminate}}\;
    }
}

\textbf{return} $M$

\end{algorithm}

This single-pass hash map lookup enables efficient matching: rather than querying each worker's hash map separately ($O(B \times W)$), we query the unified hash map once per block ($O(B)$), aggregating match results across all workers simultaneously.

\subsubsection{Sampled Prefix Hashing}

For prefix matching on worker nodes, workers use sampled prefix hashing to balance matching granularity with storage overhead. When a cached block contains fewer tokens than a threshold (e.g., 208 tokens), only that length is hashed. For larger blocks, multiple hash entries are created at regular intervals.

Specifically, for a block with $n \geq 208$ tokens, hash entries are created at positions: 208, 212, 216, 220, 224, 228, $\ldots$, up to $n$. This sampling strategy, parameterized by start threshold (208) and step size (4), enables prefix matching at multiple granularities while controlling metadata overhead.

When prefix matching occurs, the system matches against all sampled hash positions for each block. Matched blocks are categorized as either full (all tokens cached) or partially-filled (watermark indicates available space). Full blocks use reference counting for concurrent access by multiple requests, while partially-filled blocks are exclusive and allow requests to append tokens directly after the watermark.

\subsubsection{Remote KV Cache Manager Server}

The Remote KV Cache Manager Server is deployed as per-datacenter instances for capacity management and fault isolation. Unlike the hash-based Local Manager, it maintains a simple mapping structure: \textit{cache key} $\rightarrow$ \textit{file path}, optimized for persistent storage lookups on 3FS. It provides durability guarantees through persistent metadata storage, enabling cache recovery after system restarts.

\subsubsection{Cache Matching and Scheduling Integration}

When Master receives a scheduling request, it performs parallel lookups to both Local and Remote KV Cache Managers:

\begin{itemize}[leftmargin=1em]
    \item \textbf{Local Cache Query}: Query Local KV Cache Manager's unified hash map for worker-level cache matches, returning maximum match length per worker
    \item \textbf{Remote Cache Query}: Query Remote KV Cache Manager Server for 3FS cache matches, returning maximum match length from persistent storage
\end{itemize}

Both queries execute concurrently. Master combines results to compute the cache reuse score for each candidate worker $w$:
\begin{equation}
\begin{split}
score(w) = & \alpha \cdot \frac{local\_match\_len(w)}{total\_seq\_len} \\
         & + \beta \cdot \frac{remote\_match\_len}{total\_seq\_len} \\
         & - \gamma \cdot \frac{predicted\_latency(w)}{max\_latency}
\end{split}
\end{equation}
where $\alpha$, $\beta$, and $\gamma$ are weighting factors tuned based on workload characteristics. This score is combined with worker load information to make final scheduling decisions.

If the request includes a chat ID, Master uses it as a strong hint: if the chat session was previously assigned to a worker with cached KV state, Master prioritizes routing to that worker, enabling direct reuse of local cache.
\section{Speculative Decoding Design}
\label{sec:speculative_decoding}

This section presents RTP-LLM's speculative sampling framework, which enables Aone Copilot to achieve 1000 tokens per second inference performance in production deployments \cite{Aone_AoneCopilot_2025}. We examine the theoretical foundations, implementation architecture, and performance characteristics of speculative sampling in large-scale LLM deployment.

\subsection{RTP-LLM Speculative Sampling Framework}

The RTP-LLM framework implements a comprehensive speculative sampling system designed to support multiple speculative sampling algorithms while maintaining modularity and extensibility. {The framework's underlying implementation is built in C++}.

\subsubsection{Architecture Overview}

The framework decomposes speculative sampling into four modular components: i)\textbf{ProposeExecutor}: Manages token proposal generation across different algorithms (naive speculative sampling, Prompt Lookup, Eagle, MTP); ii) \textbf{ScoreExecutor}: Handles parallel token scoring by the target model; iii) \textbf{SpeculativeSampler}: Implements verification algorithms to determine token acceptance; and iv) \textbf{SpeculativeUpdater}: Updates accepted tokens to the original stream.

{The execution flow is as below: (1) \textbf{ProposeExecutor} generates $k$ candidate tokens using the configured proposal algorithm; (2) \textbf{ScoreExecutor} performs parallel forward passes through the target model to score all $k$ candidate tokens simultaneously; (3) \textbf{SpeculativeSampler} applies verification algorithms (e.g., standard speculative sampling acceptance criteria) to determine which candidate tokens to accept based on the probability distributions; (4) \textbf{SpeculativeUpdater} integrates the accepted tokens into the original generation stream, advancing the generation state accordingly. Each component maintains clear input/output interfaces and stateless operation, ensuring loose coupling and facilitating algorithm experimentation.}

\subsubsection{Supported Algorithms}

The framework supports multiple speculative sampling approaches: i) \textbf{Naive Speculative Sampling}: Direct use of smaller GPT models as propose models; ii) \textbf{MTP (Multi-Token Prediction)} \cite{xia2024unlocking, deepseekai2024deepseekv3technicalreport}: Predicts multiple next tokens in one forward pass for parallel verification (e.g., DeepSeek-V3); iii)  \textbf{Eagle} \cite{li2024eagle}: Novel Auto-Regression Head training for future hidden state prediction; iv) \textbf{Prompt Lookup}: N-gram matching against historical prompts for token proposal.

\subsection{Prompt Lookup Speculative Sampling}

Prompt Lookup represents a specialized form of speculative sampling particularly effective for extractive scenarios, where generated content can be directly copied from input prompts. The algorithm operates through n-gram token matching against the input prompt using recently generated tokens, extracting subsequent $k$ tokens as candidate proposals, and validating them through the score model. Algorithm~\ref{alg:ngram_matching_speculative} presents the complete procedure.

\begin{algorithm}[t]
\caption{\textsc{N-gram Token Matching and Speculative Sampling}}
\label{alg:ngram_matching_speculative}
\LinesNumbered
\KwIn{Input prompt $prompt$, Recently generated tokens $recent\_tokens$, Proposal count $k$}
\KwOut{Accepted tokens $accepted\_tokens$}

\tcp{Match n-grams}
$match\_result \gets$ MatchNGrams($prompt$, $recent\_tokens$)\;
% \textcolor{blue}{\Comment{Match n-grams}}\;

\If{$match\_result$ succeeds}{
    \tcp{Extract k candidates}
    $candidates \gets$ ExtractTokens($match\_result$, $k$)\;
    % \textcolor{blue}{\Comment{Extract k candidates}}\;
    \tcp{Validate}
    $validation\_results \gets$ ScoreModel($candidates$)\;
    % \textcolor{blue}{\Comment{Validate}}\;
    \tcp{Verify and accept}
    $accepted\_tokens \gets$ VerificationAlgorithm($candidates$, $validation\_results$) \;
    % \textcolor{blue}{\Comment{Verify and accept}}\;
}

\Return{$accepted\_tokens$}

\end{algorithm}
% \vspace{-4mm}
For code editing applications \cite{Aone_AoneCopilot_2025}, Prompt Lookup benefits from additional optimizations: \textbf{cursor maintenance} to track the last successful lookup position ensuring sequential copying, \textbf{skip initial matching} to use the first $k$ tokens directly in the initial iteration, and \textbf{position updates} to advance cursor position after each successful iteration. These optimizations leverage the sequential nature of code copying, where subsequent operations continue from previously copied positions.

\section{Runtime and System Extensions}
\label{sec:extension}
RTP-LLM also incorporates a series of runtime and system-level optimizations for efficient large-scale inference, including \textbf{Parallel Execution}, \textbf{Quantization}, and \textbf{Multimodal Model Support}. 

\subsection{Parallel Execution}

\label{subsec:multi-level-parallelism}
The enormous parameter count of state-of-the-art Large Language Models (LLMs), often exceeding hundreds of billions (e.g., 96B), necessitates sophisticated \textbf{multi-level parallelism strategies} to distribute computational load and memory requirements across multiple nodes and devices. Effective parallelism is critical for minimizing the Time-To-First-Token (TTFT) and maximizing system throughput in production environments. We employ a hierarchical parallelism framework integrating Tensor Parallelism, Pipeline Parallelism, Data Parallelism, and Expert Parallelism.

\begin{itemize}[leftmargin=1em]
\item \textbf{Tensor Parallelism (TP)}: TP is applied to individual weight matrices (e.g., in FFN and Attention layers) within a transformer block, partitioning them across multiple GPUs within a node. This technique addresses the single-GPU memory limit for the largest models and is crucial for accelerating the compute-bound operations (e.g., matrix multiplications) during the \textbf{Prefill phase} (context processing). High-speed interconnects like NVLink are essential for efficient inter-GPU communication during all-gather and reduce-scatter operations.
\item \textbf{Pipeline Parallelism (PP)}: PP distributes consecutive layers of the model across a sequence of GPUs, forming a pipeline. PP is essential for handling models that cannot fit even with TP on a single node. PP introduces \textbf{pipeline bubbles}, which can be mitigated by micro-batching.
% and techniques like the Gpipe and PipeDream algorithms.
\item \textbf{Data Parallelism (DP)}: DP is applied at the cluster level, replicating the model weights across multiple nodes. It is primarily used to scale throughput by processing multiple batches of requests simultaneously. This is combined with sophisticated dynamic batching and load balancing to maintain high GPU utilization.
\item \textbf{Expert Parallelism (EP)}: EP is designed for \textbf{Mixture-of-Experts (MoE)} models. The sparse FFN experts are distributed across multiple devices, and during inference, the routing mechanism only activates a small subset of experts. EP is memory-efficient and good for scaling MoE models, though it increases complexity in load balancing and dynamic routing communication.
\end{itemize}

\subsection{Quantization Techniques}
\label{sec:quantization}

Model quantization is an indispensable optimization for LLM inference, addressing the critical challenges of memory capacity and bandwidth saturation. Reducing the precision of model weights and intermediate states from FP16/BF16 to lower-bit formats (e.g., INT8/INT4) directly enhances throughput and enables the deployment of larger models on commodity hardware.

\subsubsection{Weight-Only Quantization}
\label{subsec:weight-quantization}

As Model weights constitute the majority of the model's memory footprint, we primarily employ \textbf{Weight-Only Quantization} methods. These techniques convert the weights to lower precision (typically INT4/INT8) while keeping the activations in higher precision for computation (e.g., FP16/BF16) to minimize quality loss. State-of-the-art methods include:
\begin{itemize}[leftmargin=1em]
\item \textbf{GPTQ (General Pipelined Token Quantization)} \cite{frantar2022gptq}: As an efficient, one-shot Post-Training Quantization (PTQ) method, GPTQ remains foundational for its fast implementation. However, industry focus is shifting towards more robust frameworks.
\item \textbf{AQT (Activation-aware Quantization) Frameworks}: Modern approaches such as \textbf{AWQ} (Activation-aware Weight Quantization) \cite{lin2024awq} and \textbf{HQQ} (Half-Quadratic Quantization) \cite{badri2023hqq} are preferred for pushing to extremely low bitwidths (e.g., INT3, INT2) by leveraging activation distributions or advanced non-linear optimization techniques, ensuring minimal degradation in LLM reasoning capability.
\end{itemize}
Furthermore, the emerging industry standard of \textbf{FP8 (8-bit Floating Point)} is integrated, offering high performance and minimal accuracy degradation, particularly when combined with hardware accelerators supporting the format.

\subsubsection{KV Cache Quantization}
\label{subsec:kv-cache-quantization}

The Key-Value (KV) cache, which stores intermediate attention states, dynamically grows with context length and quickly becomes the bottleneck in memory bandwidth and capacity, especially for models supporting contexts of 128K+ tokens. To mitigate this memory pressure, we employ standard quantization techniques applied specifically to the KV Cache.

\begin{itemize}[leftmargin=1em]
\item \textbf{On-the-fly Quantization}: The Key and Value tensors are quantized from FP16/BF16 to lower precision (typically INT8, INT4 or FP8) during the generation. It directly uses \textbf{per-tensor or per-block dynamic scaling} to determine the quantization factor, prioritizing hardware efficiency and speed.
\item \textbf{Memory Footprint Reduction}: Quantizing the KV cache effectively reduces its size by half (for INT8/FP8) or more. This directly alleviates the memory bandwidth pressure associated with the Decode Phase. This is essential for maximizing the effective batch size and concurrent request capacity of the engine under memory-bound conditions.
\end{itemize}
This memory-centric optimization aims to maximize concurrent request capacity and support extremely long-context workloads.

\subsection{Multimodal Model Support}
Multimodal models represent a significant advancement in artificial intelligence, enabling communication through multiple modalities with computers. These models aim to process and understand multi-modal information, including images, videos, and audio. In the context of RTP-LLM, our focus is primarily on models that accept images as input, specifically supporting prominent multimodal architectures such as LLaVA \cite{liu2023visual} and Qwen-VL \cite{bai2023qwenvl}.
\begin{figure}
  \centering
  \includegraphics[width=0.9\linewidth]{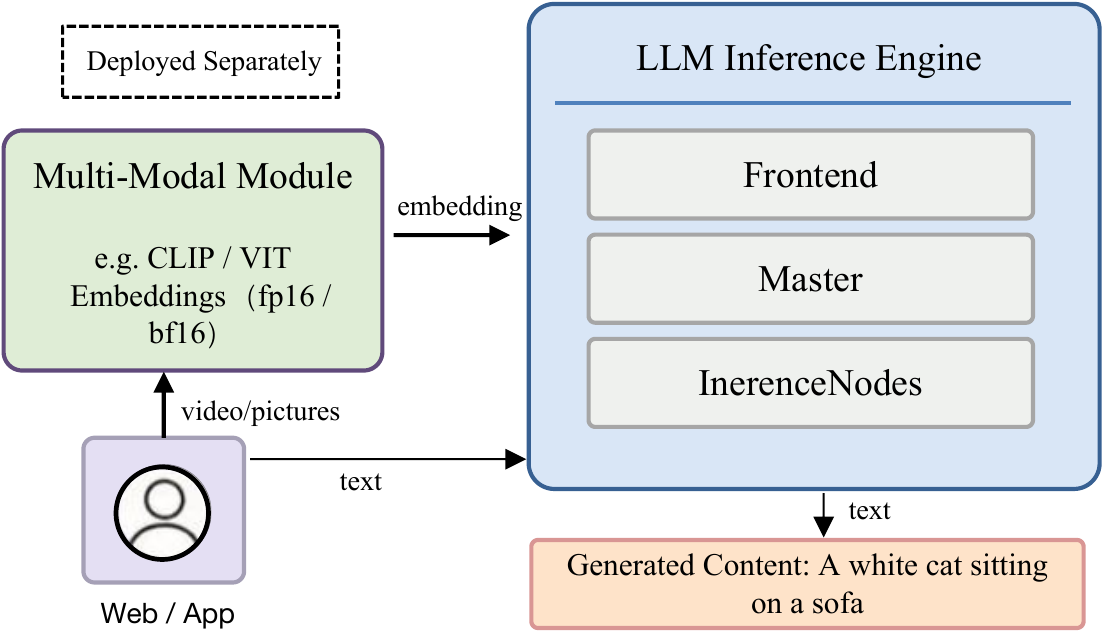}
  % \vspace{-4mm}
  \caption{EPD Disaggregation}
  \label{fig:vit_disaggration}
  % \vspace{-8mm}
\end{figure}

\subsubsection{Supported Multimodal Architectures}~
LLaVA Architecture:
The LLaVA (Large Language and Vision Assistant) model integration follows the HuggingFace format specification. The configuration file (\texttt{config.json}) contains the \texttt{mm\_vision\_tower} keyword, which specifies the path to the Vision Transformer (ViT) component \cite{dosovitskiy2020image}. Typically, this implementation utilizes OpenAI's pre-trained CLIP model for visual feature extraction.

\textbf{Invocation Interface:} The invocation mechanism maintains consistency with the HuggingFace format. Users specify image insertion positions using the \verb|<image>| tag within the prompt. Images are provided as a sequence in \texttt{List[str]} format. Notably, RTP-LLM's multimodal interface supports inserting multiple images in a single prompt, though the effectiveness of current supported models on multiple images remains limited. It is crucial to ensure strict correspondence between the number of image tags and the number of provided images.

Qwen-VL Architecture:
The Qwen-VL implementation differs slightly from LLaVA in its architectural approach. While Qwen-VL's ViT component also employs CLIP, its parameters are integrated with the Large Language Model (LLM) portion, resulting in ViT parameters being read directly from the model checkpoint rather than from separate configuration files.

\textbf{Invocation Interface:} Similar to LLaVA, Qwen-VL follows HuggingFace format conventions. Images are marked using the \verb|<img>{img_url}</img>| tag within prompts. Additionally, the system supports \verb|<img/>| placeholder syntax, enabling separation between URL specification and prompt input for enhanced flexibility.

\begin{figure*}[t]
    \setlength{\abovecaptionskip}{0cm}
    \setlength{\belowcaptionskip}{-0.4cm}
    \centering
    \footnotesize
    \stackunder[0.5pt]{\includegraphics[scale=0.3]{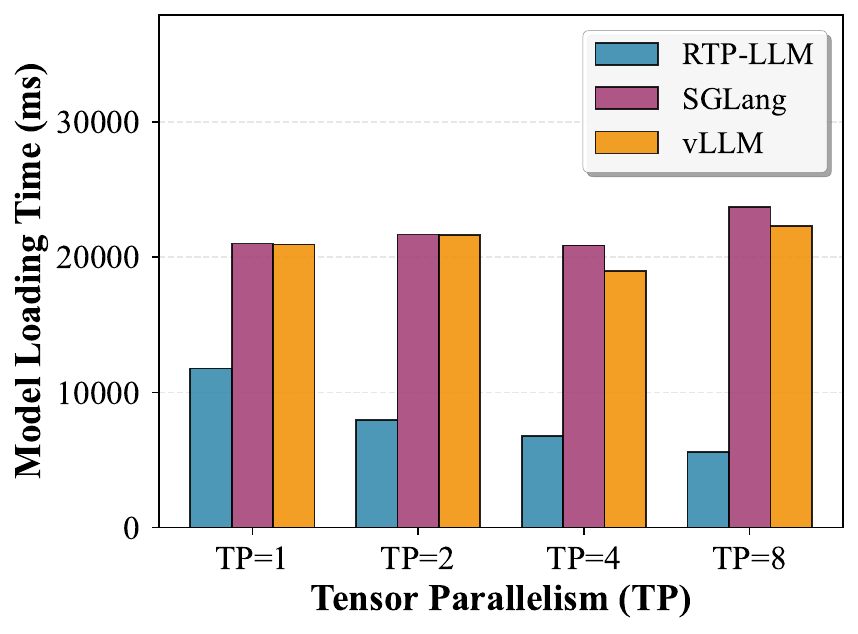}}{(a) Qwen3-32B}
    \stackunder[0.5pt]{\includegraphics[scale=0.3]{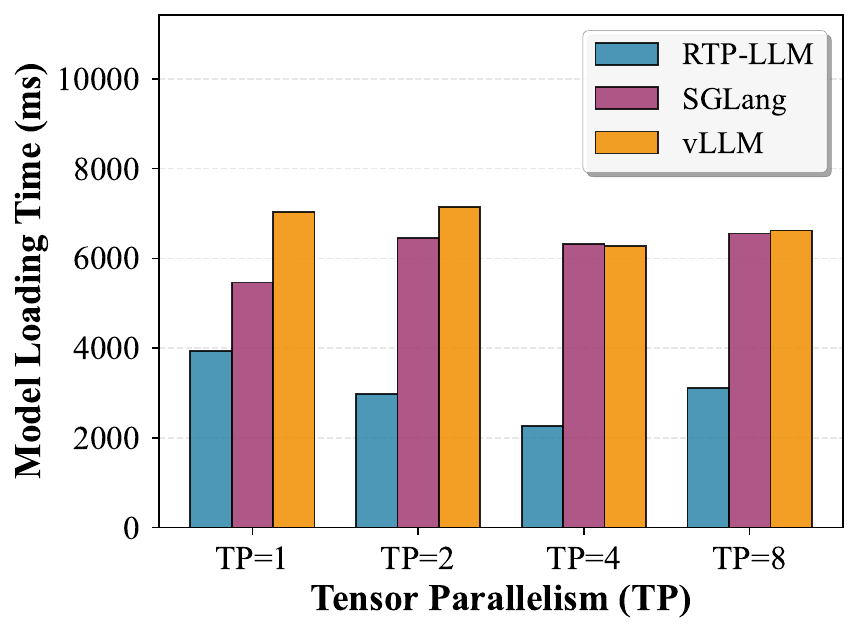}}{(b) Qwen3-8B}
    %\newline
    \stackunder[0.5pt]{\includegraphics[scale=0.3]{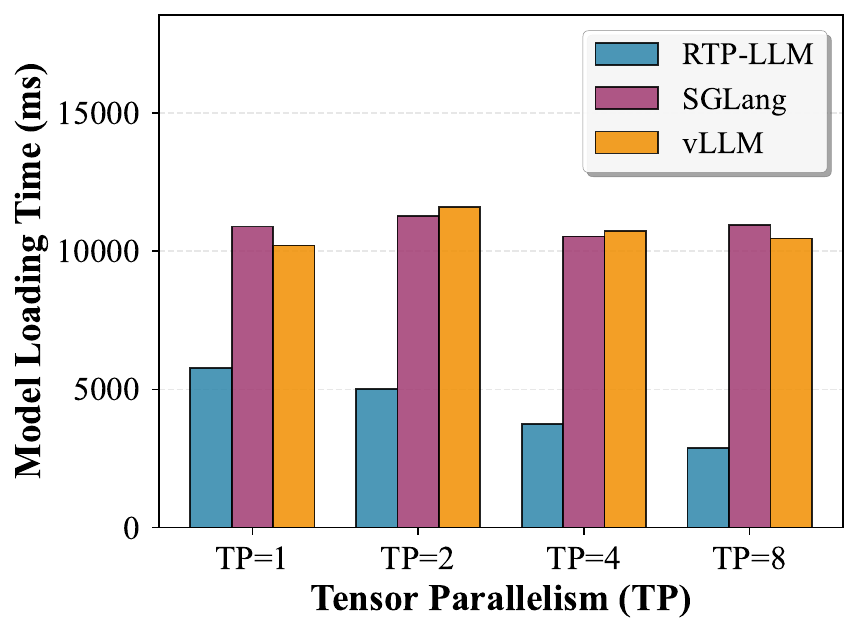}}{(c) Qwen2.5-14B-Instruct}
    \stackunder[0.5pt]{\includegraphics[scale=0.3]{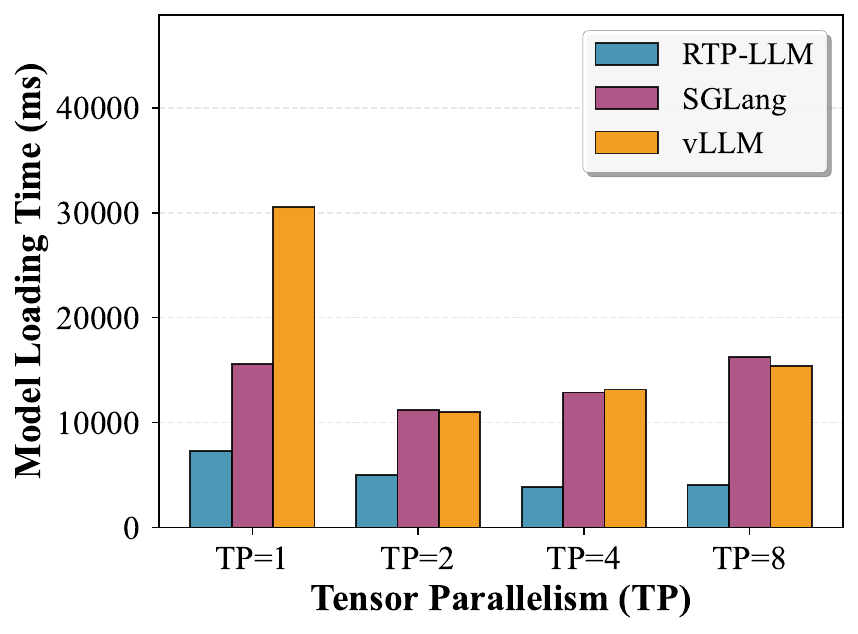}}{(d) Moonlight-16B-A3B}
    \caption{Model loading time comparison for medium-scale models (8B-32B parameters) across different TP configurations.}
    \label{fig:model_load_medium}
\end{figure*}

% \vspace{-4mm}
\subsubsection{Service Deployment}
For production use, RTP-LLM supports server-based multimodal model serving. In general, users provide the model with video or image inputs, along with corresponding textual input. The visual data (video or image) is first fed into the ViT model, which generates embedding data. This generated embedding data is then concatenated with the textual input and provided to the Language Model (LM) to ultimately produce the final output. In Figure \ref{fig:vit_disaggration}, we present the actual deployment architecture for our Vision Transformer (ViT) model. We adopted a decoupled, independent deployment strategy for the Large Language Model (LLM) and the multimodal model \cite{singh2025efficientlyservinglargemultimodal}. This architecture enables ViT models and language models to utilize separate streams during inference, avoiding contention between them. This design allows for computation overlap under high request concurrency, thereby improving overall performance. Additionally, this design provides a hidden benefit by reducing the actual GPU memory footprint on a single device.

This multimodal support framework provides a robust foundation for integrating vision-language capabilities into large-scale language model deployments, addressing the growing demand for multimodal AI applications in industrial settings.
\vspace{-2mm}
\section{Experiments}
\label{sec: eval setup}
%This section details the experimental setup, encompassing the datasets used, the specific Large Language Models (LLMs) evaluated, detailed configuration parameters, and the key performance metrics employed to quantify the system's effectiveness.
To comprehensively validate the superiority and effectiveness of the RTP-LLM inference framework, we present a set of evaluations under various LLMs and real production environments, and compare its performance against two prominent state-of-the-art open-source frameworks: \textbf{vLLM} \cite{vllm} and \textbf{SGLang} \cite{sglang}.

We {\textbf{prioritize evaluations that use real business traffic}} on traffic scheduling, PD disaggregation, and speculative decoding; then the evaluations use controlled benchmarks on model loading, quantization, and multimodal (ViT/EPD).
%use {\textbf{real production traffic from live deployment environments}};
%model loading and quantization use controlled or public settings (GQA, WikiText-2 for PPL) for reproducibility; multimodal (ViT/EPD) is evaluated both on the public GQA benchmark and with \textbf{real production traffic} from a live VL deployment.
{\textbf{For comparisons of RTP-LLM with SGLang and vLLM on production traffic}} (PD Disaggregation and Speculative Sampling), the workload is \textbf{input 200K tokens, output 16K tokens} throughout. % Each subsection explicitly states whether the data source is real business traffic or a public benchmark.
{All evaluations are conducted on servers running Linux kernel version 5.10 (x86\_64), each equipped with 64 CPU cores, 600GB memory, and 8 GPUs.}

Table~\ref{tab:evaluation_metrics} presents a summary of the core metrics employed in our evaluations to quantify the performance and industrial readiness of the RTP-LLM framework. The evaluation is structured as follows. %{\textbf{We present first the three parts that use real business traffic and live production data}} (items 1--3 below), then controlled benchmarks (items 4--6):
\begin{enumerate}[leftmargin=1em]
    \item \textbf{Production Traffic Scheduling Effectiveness (addressing Challenge I and II):} Validating the impact and stability of traffic scheduling with KV cache management under \textbf{real deployment environments}, demonstrating GPU utilization and memory efficiency in production.
    \item \textbf{Deployment Evaluation of PD Disaggregation (addressing Challenge I and III):} Evaluating Prefill-Decode Disaggregation with large-scale MoE models using \textbf{real business traffic} from actual online deployments, demonstrating GPU utilization and architectural flexibility for heterogeneous model types.
    \item \textbf{Speculative Decoding Performance Gain (addressing Challenge I):} Quantifying the acceleration and throughput improvement. We further report one \textbf{real deployment scenario} for {\textbf{user code editing prompt generation}}.
    %: (a)~235B MoE with MTP under 1K real merchant data-agent queries (input context $\sim$20K tokens; avg.\ input 19.5K, output 800). %and (b)~480B MoE MTP with input context lengths of 30K and 62K tokens, with analysis of online vs.\ offline parallelism choices.
    \item \textbf{Model Loading Superiority (addressing Challenge IV):} Benchmarking loading latency against SOTA baselines to confirm superiority for rapid elastic scaling and model iteration.
    \item \textbf{Quantized Inference Efficiency (addressing Challenge II):} Assessing throughput and memory footprint improvements from advanced quantization and KV cache management, demonstrating memory efficiency gains.
    \item \textbf{Multimodal Inference Performance (addressing Challenge III):} Evaluating end-to-end efficiency of complex multimodal models via decoupled ViT and LLM (EPD) processing: the public GQA benchmark for framework comparison, and \textbf{real production deployment} %(VL 235B, real online query, input 4K, avg.\ output 1,600 tokens)
    for RTP-LLM RT and TTFT across concurrency, demonstrating architectural flexibility in production.
\end{enumerate}

\vspace{-3mm}
\subsection{Traffic Scheduling Strategies Validation}

{We use real deployment environments within Alibaba: \textbf{internal robot Q\&A service} and \textbf{Taobao merchant customer service consultation}. The reported metrics are TTFT P95, inference P95, and cache reuse length. We evaluate RTP-LLM's traffic scheduling with KV cache management, comparing against baseline strategies without traffic scheduling.}

{Table~\ref{tab:traffic_scheduling_performance} presents the latency comparison.} For {\textbf{internal robot Q\&A service}} (Qwen 7B model, input length 300--1000 tokens, average 340 tokens, output length 6--7 tokens), RTP-LLM with traffic scheduling achieves significant improvements: TTFT P95 latency reduces from 83.3\,ms to 52.3\,ms (37.2\% reduction), while maintaining similar average inference latency; the improved cache reuse (Table~\ref{tab:kv_cache_reuse}) enables 75\% reduction in prefill machine count (from 80 to 20 machines) while maintaining average TTFT performance.
For {\textbf{Taobao merchant customer service consultation}} (Qwen 4B model, input length 2500 tokens, output length 114 tokens), RTP-LLM with traffic scheduling also demonstrates consistent performance improvements compared to that without traffic scheduling. RTP-LLM reduces TTFT P95 latency from 350\,ms to 226\,ms (35.4\% reduction), and inference P95 latency from 1760\,ms to 1210\,ms (31.3\% reduction).

\begin{table}[tb]
    \centering
    \caption{Key metrics for performance evaluation} \vspace{-4mm}
    \label{tab:evaluation_metrics}
    \begin{tabular}{lccp{3.5cm}}
        \toprule
        \textbf{Metric} & \textbf{Unit} & \textbf{Description} \\
        \midrule
        TTFT & ms & Time to first token \\
        Tokens/s & T/s & System throughput  \\
        GPU Memory & MB & Peak Runtime Memory Usage \\
        Cache Hit Rate & \% & Efficiency of KV cache management \\
        Batch Latency & ms & Total cost time for batch execution \\
        PPL & N/A & Model output fidelity  \\
        \bottomrule
        \vspace{-4mm}
    \end{tabular}
\end{table}

\begin{table}[t]
    \centering
    \caption{Traffic scheduling performance comparison across two real production workloads. TS: Traffic Scheduling. Latency is measured in milliseconds (ms). \vspace{-4mm}}
    \label{tab:traffic_scheduling_performance}
    \begin{tabular}{lccc}
        \toprule
        \textbf{Workload} & \textbf{TS} & \textbf{TTFT P95} & \textbf{Inference P95} \\
        \midrule
        \multirow{2}{*}{Internal Robot Q\&A} & Off & 83.3 & 136 \\
        & On & 52.3 & 96.2 \\
        \midrule
        \multirow{2}{*}{Taobao Merchant Service} & Off & 350 & 1760 \\
        & On & 226 & 1210 \\
        \bottomrule
    \end{tabular}
\vspace{-2mm}
\end{table}

\begin{table}[t]
    \centering
    \vspace{-2mm}
    \caption{KV cache reuse length comparison across different strategies. Cache reuse length is measured in tokens.}\vspace{-4mm}
    \label{tab:kv_cache_reuse}
    \begin{tabular}{lcc}
        \toprule
        \textbf{Workload} & \textbf{TS} & \textbf{Cache Reuse Length} \\
        \midrule
        \multirow{2}{*}{Internal Robot Q\&A} & Off & 26.6 \\
        & On & 83.8 \\
        \midrule
        \multirow{2}{*}{Taobao Merchant Service} & Off & 833 \\
        & On & 840 \\
        \bottomrule
    \end{tabular}
   % \vspace{2mm}
 % \vspace{-8mm}
\end{table}

{Table~\ref{tab:kv_cache_reuse} summarizes cache reuse. RTP-LLM's unified hash map and cache affinity-based routing improve reuse length: for {internal robot Q\&A}, from 26.6 to 83.8 tokens (215\% improvement); for {Taobao merchant service}, from 833 to 840 tokens. This reduces prefill overhead and latency.}

\subsection{PD Disaggregation Performance}
%{\textbf{We use real business traffic from actual online deployments}} of the Qwen3-Coder-480B-FP8 model; the workload and batch settings match production traffic patterns.
We evaluate RTP-LLM's Prefill-Decode Disaggregation architecture using the Qwen3-Coder-480B-FP8 model \cite{qwen3technicalreport}, a large-scale MoE (Mixture-of-Experts) model with FP8 quantization.
%and the system's capability to handle ultra-large models in production through a disaggregated serving architecture.
Note that, Qwen3-Coder-480B-FP8 is deployed in online business environment. The workload is configured with a batch size of 64 to match production traffic patterns.

\textbf{Deployment Configuration:} The deployment employs a distributed setup across 5 nodes, each equipped with 8 GPUs, implementing the Prefill-Decode Disaggregation architecture. Specifically, 4 nodes are dedicated to Prefill processing, while 1 node handles Decode operations. This asymmetric allocation reflects the typical workload characteristics where prefill requires more computational resources due to parallel processing of input sequences, while decode benefits from higher concurrency with lower per-request computational intensity.

\textbf{Parallelism and Communication:} Each node is configured with Tensor Parallelism (TP=8), Expert Parallelism (EP=8), and Data Parallelism (DP=1). The Expert Parallelism configuration uses DeepEP for efficient All2All communication among expert layers, optimizing the communication overhead in MoE models \cite{deepep2025}. For KV cache transmission between Prefill and Decode nodes in the disaggregated architecture, the system utilizes NCCL IBRC (InfiniBand Reliable Connection) \cite{NVIDIA_NCCL_UserGuide_2025} for high-performance, low-latency data transfer, ensuring efficient communication of cached key-value states across the distributed deployment. Prefill nodes operate in normal mode, prioritizing throughput for batch processing, while the Decode node operates in low latency mode, minimizing inter-token latency for real-time generation.

\textbf{Concurrency Settings:} To maximize resource utilization while maintaining latency requirements, Prefill nodes are configured with a batch size of 64, allowing efficient parallel processing of input sequences. The Decode node supports a higher concurrency of 128, capitalizing on the memory-bound nature of decode operations to process multiple requests simultaneously.

%\textbf{Data and parameters:} {\textbf{As noted above, the evaluation uses real production traffic traces from our online business deployments}}; the workload is configured with a batch size of 64 to match production traffic patterns.

\begin{table}[t]
    \centering
    \caption{Performance comparison for Qwen3-Coder-480B-A35B-Instruct-FP8 (KV Cache FP8 enabled) across different frameworks. (\textbf{RTP-LLM speedup:} Cache Hit Rate: SGLang 1.57x, vLLM 2.36x; TTFT: SGLang 4.72x, vLLM 5.33x)} \vspace{-4mm}
    \label{tab:pd_disaggregation_performance}
    \begin{tabular}{lccc}
        \toprule
        \textbf{Metric} & \textbf{RTP-LLM} & \textbf{SGLang} & \textbf{vLLM} \\
        \midrule
        Cache Hit Rate (\%) & 45.09 & 28.70 & 19.10 \\
        TTFT (ms) & 1338.38 & 6322.7 & 7134.8 \\
        Tokens/s & 1081.72 & 1152.95 & 1084.58 \\
        \bottomrule
    \end{tabular}
    %\vspace{5mm}
     % \vspace{-6mm}
\end{table}

Table~\ref{tab:pd_disaggregation_performance} presents the performance comparison for the Qwen3-Coder-480B-A35B-Instruct-FP8 model with KV Cache FP8 enabled. RTP-LLM achieves a cache hit rate of 45.09\%, outperforming SGLang (28.70\%) and vLLM (19.10\%) by 1.57x and 2.36x, respectively. RTP-LLM achieves the lowest TTFT at 1338.38 ms, demonstrating 4.72x and 5.33x speedup compared to SGLang (6322.7 ms) and vLLM (7134.8 ms). The latency improvement is primarily attributed to traffic scheduling strategies that enable more effective prefix cache reuse. Throughput is essentially consistent across all frameworks: RTP-LLM achieves 1081.72 tokens/s, compared to SGLang (1152.95 tokens/s) and vLLM (1084.58 tokens/s).

\vspace{-0.1in}

\subsection{Speculative Sampling Performance}
 %{\textbf{This part uses real business traffic}}---production request traces collected from our deployment scenarios---{\textbf{so that throughput and latency reflect actual serving conditions in production, not synthetic benchmarks or load tests.}}
 We evaluate RTP-LLM's speculative decoding performance using the DeepSeek-V3-0324 \cite{deepseekai2024deepseekv3technicalreport} model, where the last layer of the model serves as the draft model for speculative decoding. All runs use Tensor Parallelism (TP=8, DP=1) deployment and consistent parameters: max\_batch\_size = 32, max\_new\_tokens = 500, FP8 KV Cache is enabled, and the speculative decoding step size is set to 1 (predicting one token per step).

\begin{table}[t]
%\vspace{2mm}
    \centering
    \caption{Throughput comparison (tokens/s) for DeepSeek-V3-0324 speculative sampling across different frameworks. (\textbf{RTP-LLM Speedup:} compared with vLLM: 1.12x; compared with SGLang: 2.48x)} \vspace{-3mm}
    \label{tab:speculative_tokens_per_sec}
    \begin{tabular}{lccc}
        \toprule
        \textbf{Framework} & \textbf{RTP-LLM} & \textbf{vLLM} & \textbf{SGLang} \\
        \midrule
        Tokens/s & 187.53 & 167.95 & 75.785 \\
        \bottomrule
    \end{tabular}
   %\vspace{9mm}
\end{table}

Table~\ref{tab:speculative_tokens_per_sec} presents the throughput comparison in tokens per second across the three frameworks. RTP-LLM achieves the highest throughput at 187.53 tokens/s, demonstrating 1.12x speedup compared to VLLM (167.95 tokens/s) and 2.48x speedup compared to SGLang (75.785 tokens/s). The modest superior performance over vLLM stems from RTP-LLM's direct C++ operator launch mechanism, which eliminates the Python-to-C++ invocation overhead present in open-source frameworks like vLLM that require Python calls to C++ before launching operators, reducing per-operator invocation overhead in the speculative sampling pipeline.

\textbf{{Real Deployment: 235B MoE with MTP (Merchant Data-Agent)}}:
%The deployment serves {\textbf{user code editing prompt generation}} (code completion and in-editor suggestions for developers). {\textbf{Metrics are from live production; 1K real production queries}} (see Table~\ref{tab:realdeploy_235b_mtp}). %{\textbf{This part is based on production data from the real deployment, not synthetic or stress testing.}}
To validate speculative decoding under {\textbf{real production deployment}}, we collected metrics using \textbf{1,000 real production queries} from an online merchant data-agent service powered by the Qwen3-235B-A22B MoE model \cite{qwen3technicalreport}, with Multi-Token Prediction (MTP) enabled. The workload reflects actual business traffic: \textbf{input context cap 20K tokens} (average input length 19.5K tokens, average output length 800 tokens), Prefill 4TP$\times$4 rows with prefix cache reuse, Decode with the same total GPU count. The four decode configurations in Table~\ref{tab:realdeploy_235b_mtp}---4TP$\times$4, 1TP8DP, 2TP4DP, and 2TP8DP---are \textbf{all used in real production} for this service; we compare them under the same total GPU budget to guide production choice. Table~\ref{tab:realdeploy_235b_mtp} reports decode-side single-card TPS and average time-per-output-token (TPOT) at 64--512 client concurrency.

\begin{figure*}[t]
    \setlength{\abovecaptionskip}{0cm}
    \setlength{\belowcaptionskip}{-0.1cm}
    \centering
    \footnotesize
    \stackunder[0.5pt]{\includegraphics[scale=0.3]{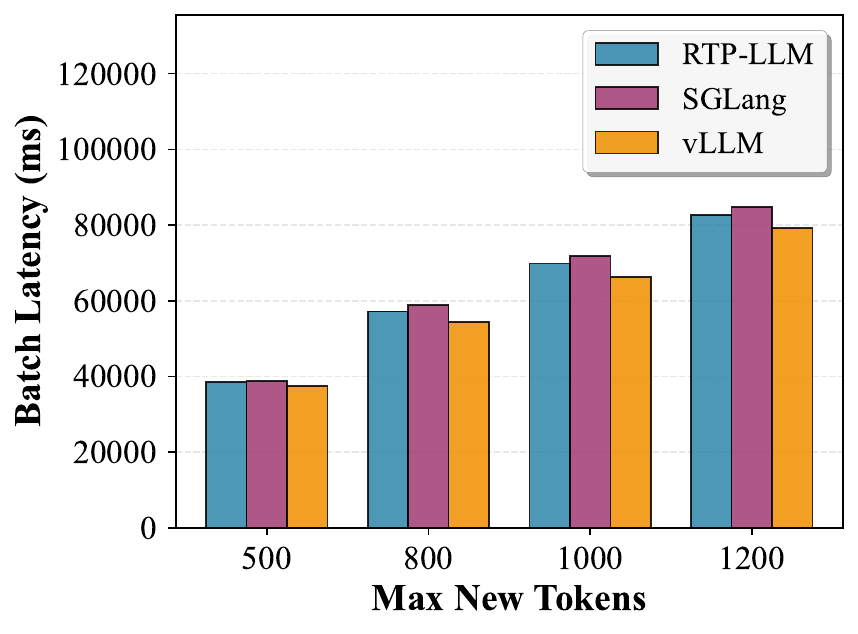}}{(a) Batch Latency (AWQ(FP8))}
    \stackunder[0.5pt]{\includegraphics[scale=0.3]{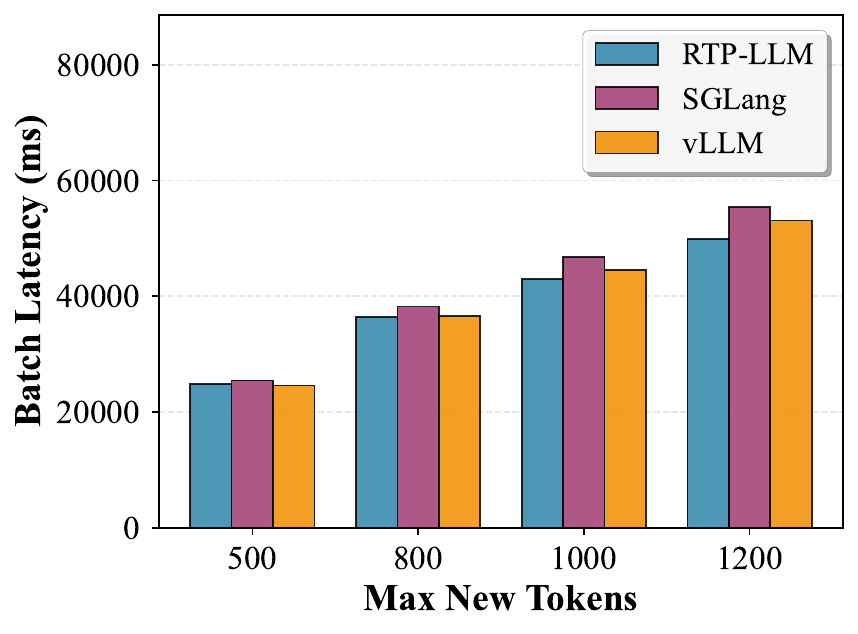}}{(b) Batch Latency (FP8 KV Cache)}
    %\newline
    \stackunder[0.5pt]{\includegraphics[scale=0.3]{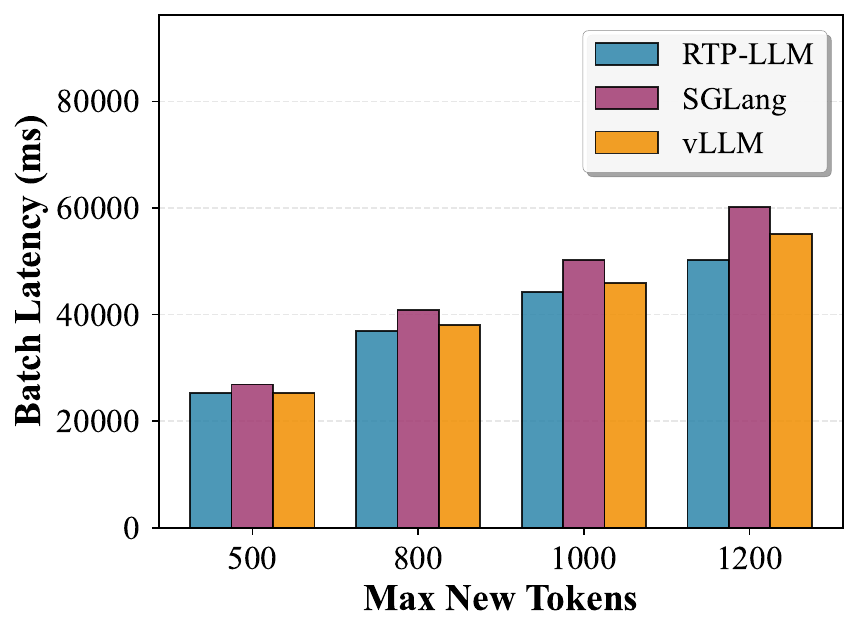}}{(c) Batch Latency (Baseline)}
    \stackunder[0.5pt]{\includegraphics[scale=0.3]{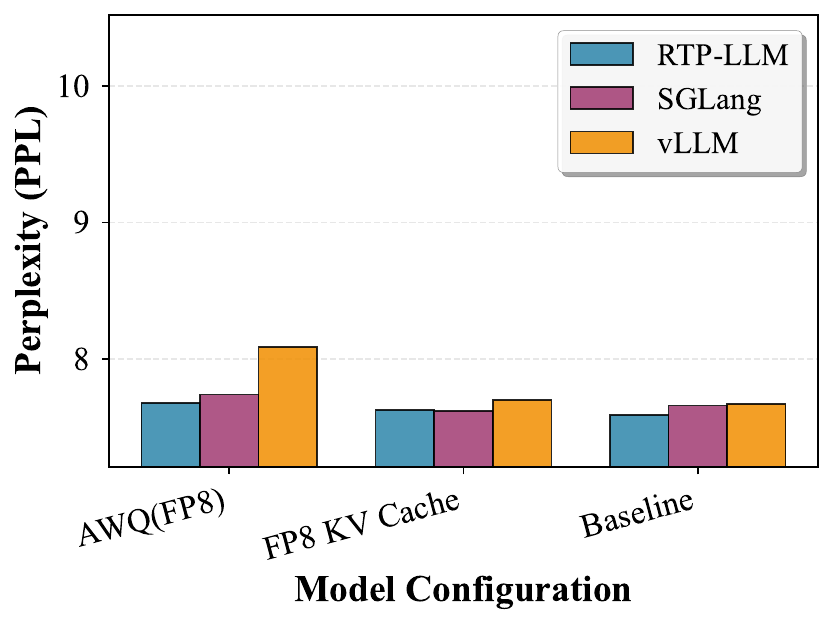}}{(d) Precision Loss (PPL)}
    \caption{Batch latency and precision loss comparison for Qwen3-32B across different quantization configurations.}
    \label{fig:quantization_batch_latency}
    \vspace{-4mm}
\end{figure*}

\textbf{Deployment Scheme Guidance.} In production, teams choose among these schemes according to whether the workload is latency-sensitive (e.g., online interactive) or throughput-oriented (e.g., offline batch). \textit{4TP$\times$4} favors \textbf{low concurrency}: it achieves the best TPOT at 64 concurrency (15.83\,ms) but TPOT and stability degrade as concurrency grows, and is typical for dedicated low-latency lanes. \textit{1TP8DP} is best for \textbf{online, latency-sensitive} serving: it consistently delivers the best TPOT at 128--512 concurrency (17.57--25.64\,ms), as higher data parallelism spreads requests and reduces per-request wait; it is the preferred option in our production environment when the goal is to minimize time-per-output-token under real traffic. \textit{2TP4DP} is best for \textbf{offline, throughput-oriented} workloads: it achieves the highest decode TPS (e.g., 601.98 at peak) by balancing tensor and data parallelism, maximizing utilization when KV cache is saturated. \textit{2TP8DP} uses more tensor parallelism and exhibits the worst TPOT at high concurrency (e.g., 46.3\,ms at 512), as cross-GPU communication under TP becomes the bottleneck; it is still a valid production option when the same cluster must serve both prefill-heavy and decode-heavy phases with a unified TP. Under this real workload, MTP maintains an effective sampling rate of approximately 1.9 tokens per step, with KV cache utilization above 90\% and SM utilization around 60\%. These results show that RTP-LLM's speculative decoding and the above parallelism choices are validated in \textbf{real production environments} where these deployment schemes are in active use.

\begin{table}[t]
    \centering
    \caption{Real deployment (235B MoE + MTP): decode TPS (single card) and average TPOT (ms). %Merchant data-agent service; 1K real queries, input $\sim$20K tokens (avg.\ 19.5K), output 800.
    }
    \label{tab:realdeploy_235b_mtp}
    \vspace{-2mm}
    \begin{tabular}{lcccc}
        \toprule
        \textbf{Decode config} & \textbf{64 conc.} & \textbf{128 conc.} & \textbf{256 conc.} & \textbf{512 conc.} \\
        \midrule
        \multicolumn{5}{l}{\textit{Single-card Decode TPS}} \\
        4TP$\times$4 & 147.9 & 270.9 & 376.7 & 427.9 \\
        1TP8DP & 86.0 & 202.1 & 380.8 & 529.8 \\
        2TP4DP & \textbf{232.9} & \textbf{381.9} & \textbf{550.1} & \textbf{599.8} \\
        2TP8DP & 143.8 & 262.8 & 421.0 & 454.4 \\
        \midrule
        \multicolumn{5}{l}{\textit{Avg.\ TPOT (ms)}} \\
        4TP$\times$4 & \textbf{15.83} & 21.54 & 27.12 & 28.4 \\
        1TP8DP & 17.87 & \textbf{17.57} & \textbf{21.81} & \textbf{25.64} \\
        2TP4DP & 16.14 & 19.45 & 27.07 & 28.45 \\
        2TP8DP & 17.21 & 22.68 & 35.8 & 46.3 \\
        \bottomrule
    \end{tabular}
    \vspace{-4mm}
\end{table}

\begin{table}[t]
    \centering
    \caption{Model loading time comparison for Qwen3-235B-A22B across different TP configurations. Loading times are measured in seconds (s). (\textbf{RTP-LLM Speedup compared with SGLang and vLLM:} TP=4: 4.70x--4.78x; TP=8: 6.18x--6.27x)} \vspace{-2mm}
    \label{tab:model_load_235b}
    \begin{tabular}{lccc}
        \toprule
        \textbf{TP} & \textbf{RTP-LLM} & \textbf{SGLang} & \textbf{vLLM} \\
        \midrule
        TP=4 & 37.1s & 177.4s & 174.3s \\
        TP=8 & 33.0s & 206.7s & 204.0s \\
        \bottomrule
    \end{tabular}
     \vspace{-1mm}
\end{table}

%\subsection{Real Deployment: 480B MoE with MTP}
%\textbf{Scenario and data source.} As in Section~8.4, the scenario is {\textbf{user code editing prompt generation}} (code completion and in-editor suggestions). {\textbf{Live deployment;}} input context 30K--62K tokens, up to 2K generated (production TP/DP configs). {\textbf{This part is based on production data from the real deployment, not synthetic testing.}}

%We further validated Multi-Token Prediction on the Qwen3-Coder-480B MoE model in real deployment configurations for this {\textbf{code editing prompt generation}} workload. In this live deployment, we used \textbf{input context lengths of 30K and 62K tokens} (with up to 2K generated tokens per request), across multiple TP/DP configurations (e.g., 8TP1DP, 2DP8TP, 4DP4TP, 8DP2TP) that correspond to \textbf{deployment schemes used in production} for large MoE models. The same trade-off applies: higher DP favors latency-sensitive serving, while balanced TP/DP favors peak throughput when the system is saturated. Under a typical accept rate of 0.85, the effective accepted tokens per step (steps 1--4) reach approximately 1.85--3.71, demonstrating that MTP delivers substantial throughput gain at these production-relevant input scales. Together with the 235B merchant scenario above, these results show that RTP-LLM's speculative decoding is effective in \textbf{real production environments} where diverse deployment schemes (varying TP/DP) are actively used to serve different business objectives.

\subsection{Model Loading Performance}

We evaluate RTP-LLM's model loading performance across five LLM models (8B-235B parameters) \cite{qwen3technicalreport, qwen2.5, qwen2, liu2025muonscalablellmtraining} under different Tensor Parallelism (TP) configurations, comparing against SGLang and vLLM.
Figure~\ref{fig:model_load_medium} shows that RTP-LLM achieves 1.4x-4.2x speedup over baseline frameworks for medium-scale models (8B-32B parameters), with performance improving at higher TP configurations. For the large-scale Qwen3-235B-A22B model \cite{qwen3technicalreport} (shown in Table~\ref{tab:model_load_235b}), RTP-LLM achieves 4.7x-6.3x speedup, reducing loading time from 37.1s (TP=4) to 33.0s (TP=8). {This superior performance stems from shared weight loading, where each GPU reads its assigned model partition in parallel, enabling higher TP configurations to achieve more speedup for large-scale models by distributing I/O load. Beyond parallel shared reading, RTP-LLM further optimizes performance by overlapping weight loading with data broadcasting, allowing each GPU to broadcast previously loaded data while simultaneously loading its assigned partition, a capability absent in baseline frameworks}. In contrast, SGLang and vLLM exhibit negative scalability, with loading time increasing by 16.5\% and 17.0\% respectively when scaling from TP=4 to TP=8. The performance advantage becomes more pronounced for larger models and higher TP configurations, indicating superior parallel loading efficiency.

\begin{figure}[t]
    \setlength{\abovecaptionskip}{0cm}
    \setlength{\belowcaptionskip}{-0.1cm}
    \centering
    \footnotesize
    \stackunder[0.5pt]{\includegraphics[scale=0.27]{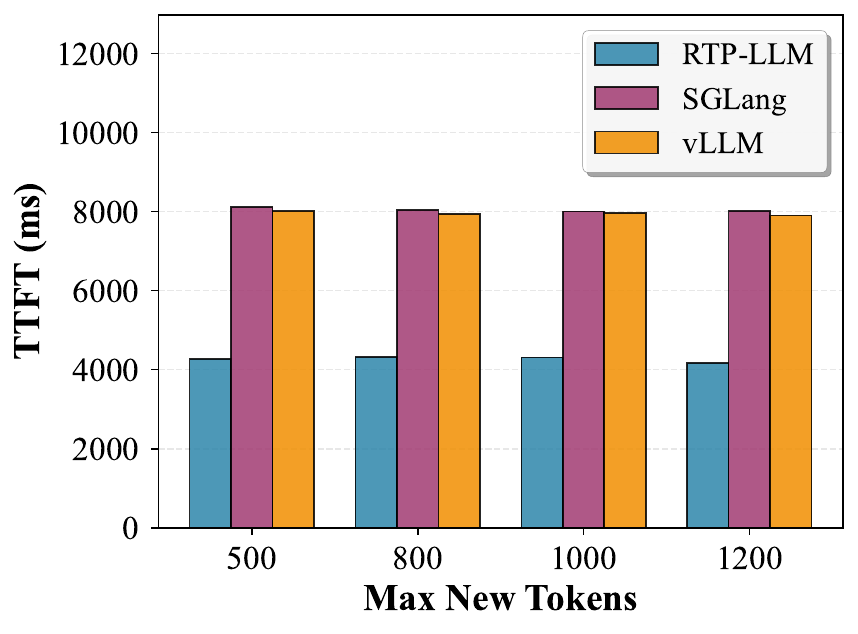}}{(a) TTFT (AWQ(FP8))}
    \stackunder[0.5pt]{\includegraphics[scale=0.27]{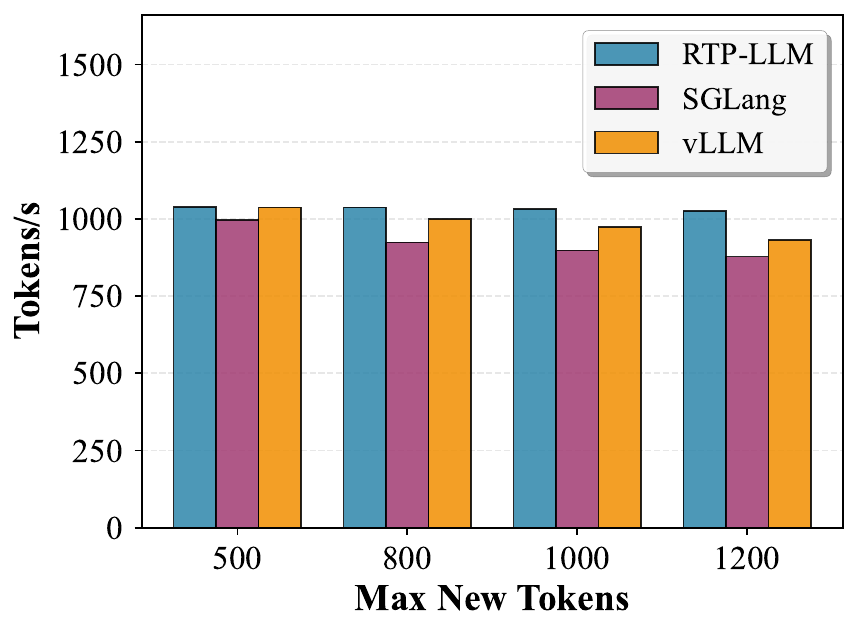}}{(b) Tokens/s (AWQ(FP8))}\\

    \stackunder[0.5pt]{\includegraphics[scale=0.27]{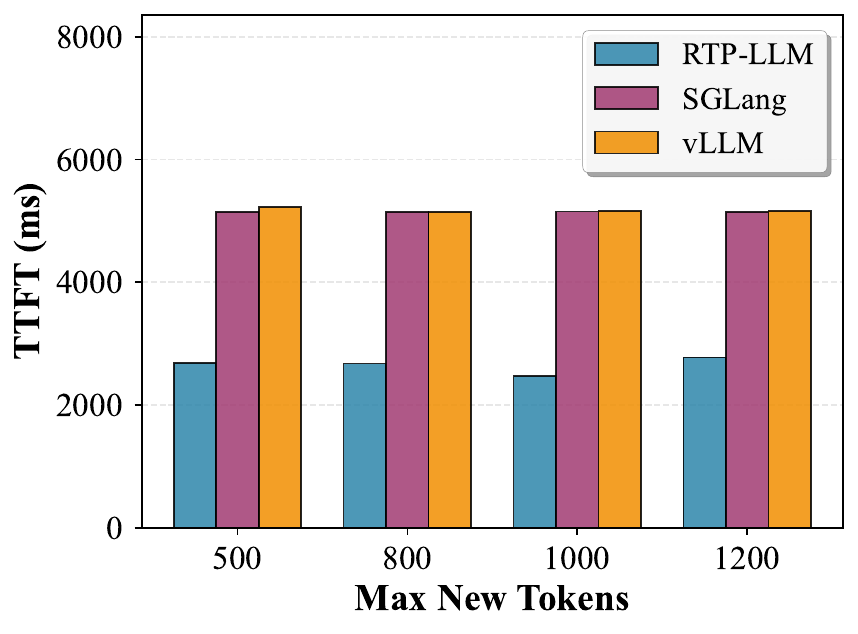}}{(c) TTFT (FP8 KV Cache)}
    \stackunder[0.5pt]{\includegraphics[scale=0.27]{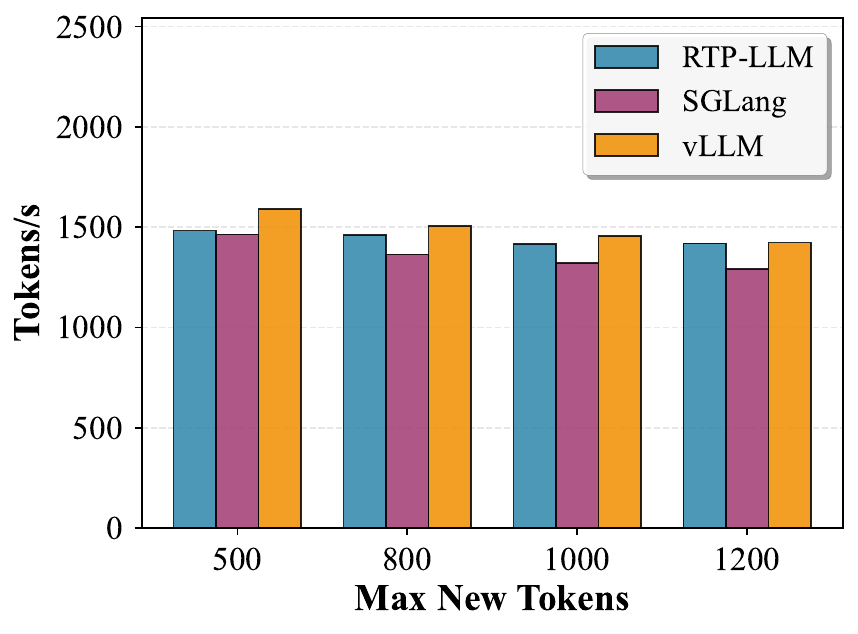}}{(d) Tokens/s (FP8 KV Cache)}
    \\
    %\newline
    \stackunder[0.5pt]{\includegraphics[scale=0.27]{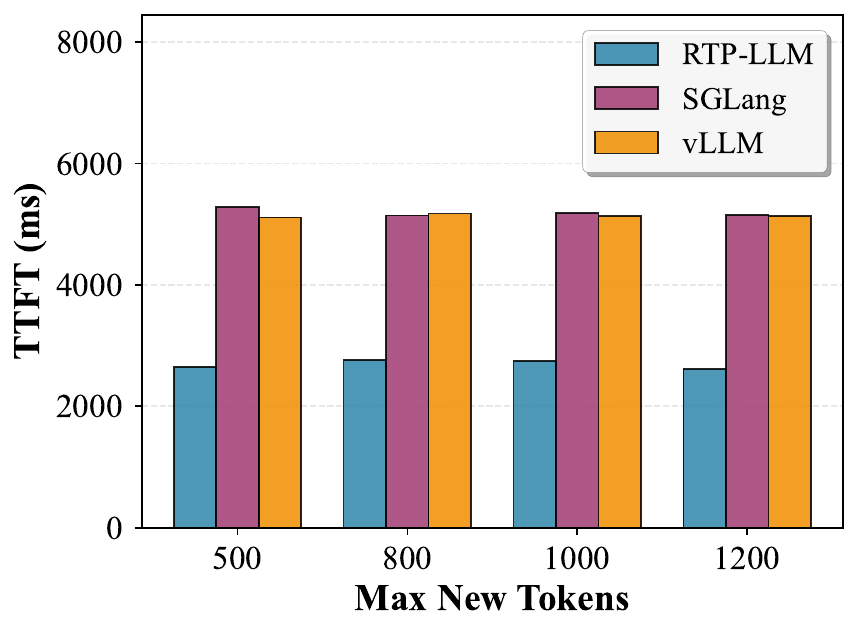}}{(e) TTFT (Baseline)}
    \stackunder[0.5pt]{\includegraphics[scale=0.27]{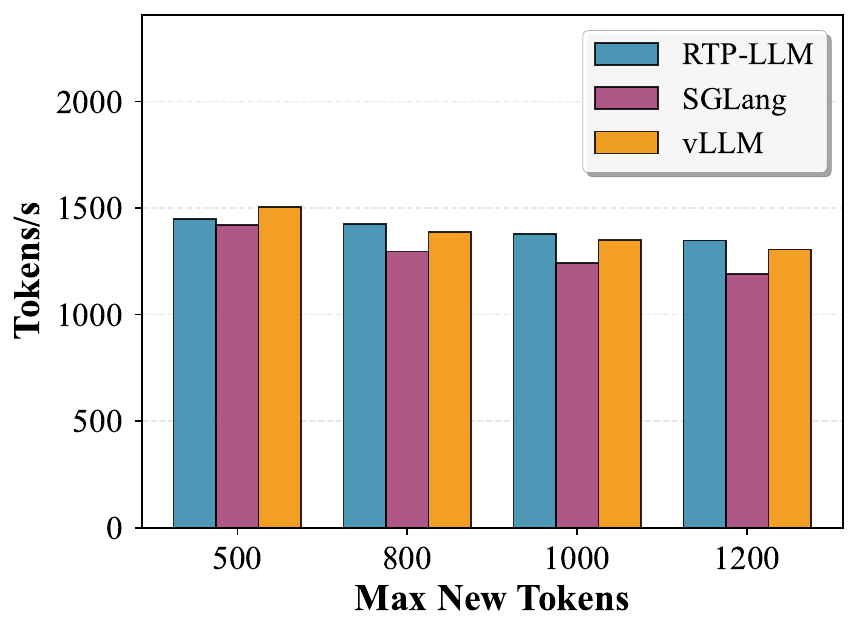}}{(f) Tokens/s (Baseline)}
    \caption{TTFT  and Tokens/s comparison for Qwen3-32B across different quantization configurations.}
    \label{fig:quantization_ttft}
\end{figure}

%\begin{figure*}[t]
%    \setlength{\abovecaptionskip}{0cm}
%    \setlength{\belowcaptionskip}{-0.1cm}
%    \centering
%    \footnotesize
 %   \stackunder[0.5pt]{\includegraphics[scale=0.3]{figures/quantization_exp/quantization_ttft_AWQFP8.pdf}}{(a) TTFT (AWQ(FP8))}
 %   \stackunder[0.5pt]{\includegraphics[scale=0.3]{figures/quantization_exp/quantization_ttft_FP8_KV_Cache.pdf}}{(b) TTFT (FP8 KV Cache)}
    %\newline
 %   \stackunder[0.5pt]{\includegraphics[scale=0.3]{figures/quantization_exp/quantization_ttft_Baseline.pdf}}{(c) TTFT (Baseline)}
%    \caption{TTFT comparison for Qwen3-32B across different quantization configurations.}
 %   \label{fig:quantization_ttft}
%\end{figure*}

%\begin{figure*}[t]
%    \setlength{\abovecaptionskip}{0cm}
 %   \setlength{\belowcaptionskip}{-0.3cm}
%    \centering
%    \footnotesize
 %   \stackunder[0.5pt]{\includegraphics[scale=0.3]{figures/quantization_exp/quantization_tokens_per_sec_AWQFP8.pdf}}{(a) Tokens/s (AWQ(FP8))}
%    \stackunder[0.5pt]{\includegraphics[scale=0.3]{figures/quantization_exp/quantization_tokens_per_sec_FP8_KV_Cache.pdf}}{(b) Tokens/s (FP8 KV Cache)}
    %\newline
%   \stackunder[0.5pt]{\includegraphics[scale=0.3]{figures/quantization_exp/quantization_tokens_per_sec_Baseline.pdf}}{(c) Tokens/s (Baseline)}
 %   \caption{Tokens/s comparison for Qwen3-32B across different quantization configurations.}
 %   \label{fig:quantization_tokens_per_sec}
   % \vspace{-2mm}
%\end{figure*}

\begin{figure*}[t]
    \setlength{\abovecaptionskip}{0cm}
    \setlength{\belowcaptionskip}{-0.2cm}
    \centering
    \footnotesize
    \stackunder[0.5pt]{\includegraphics[scale=0.3]{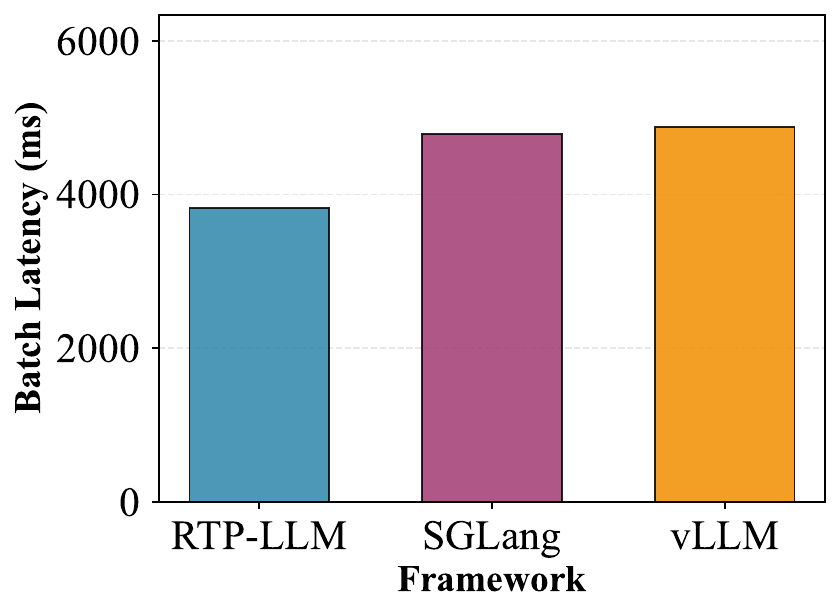}}{(a) Batch Latency}
    \stackunder[0.5pt]{\includegraphics[scale=0.3]{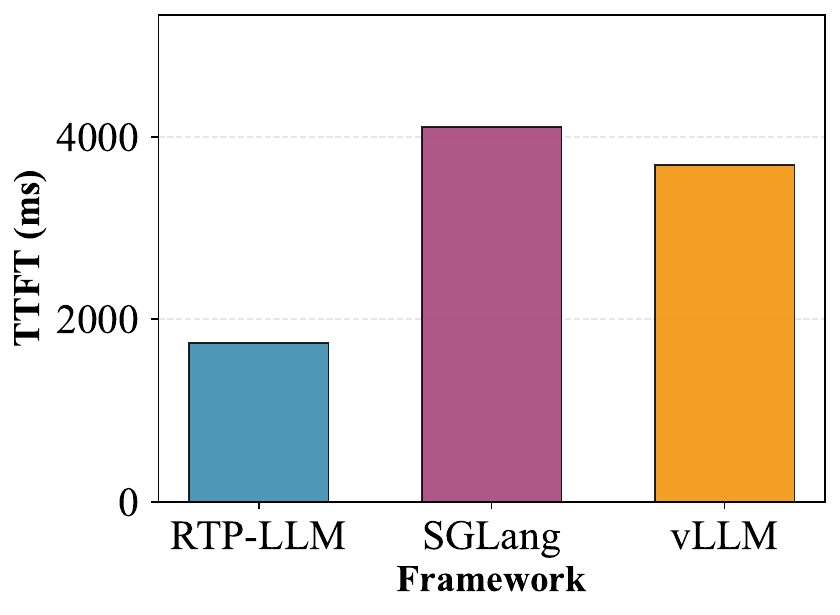}}{(b) TTFT}
    %\newline
    \stackunder[0.5pt]{\includegraphics[scale=0.3]{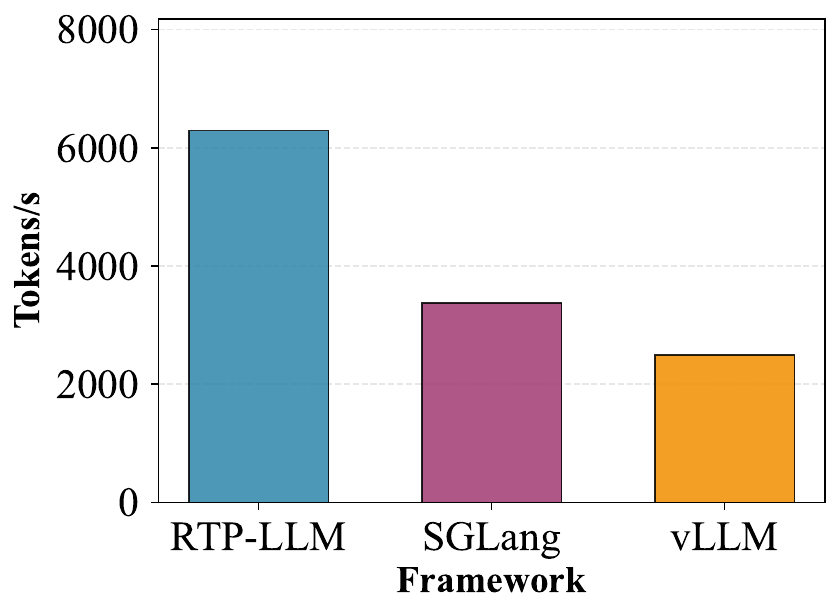}}{(c) Tokens/s}
    \stackunder[0.5pt]{\includegraphics[scale=0.3]{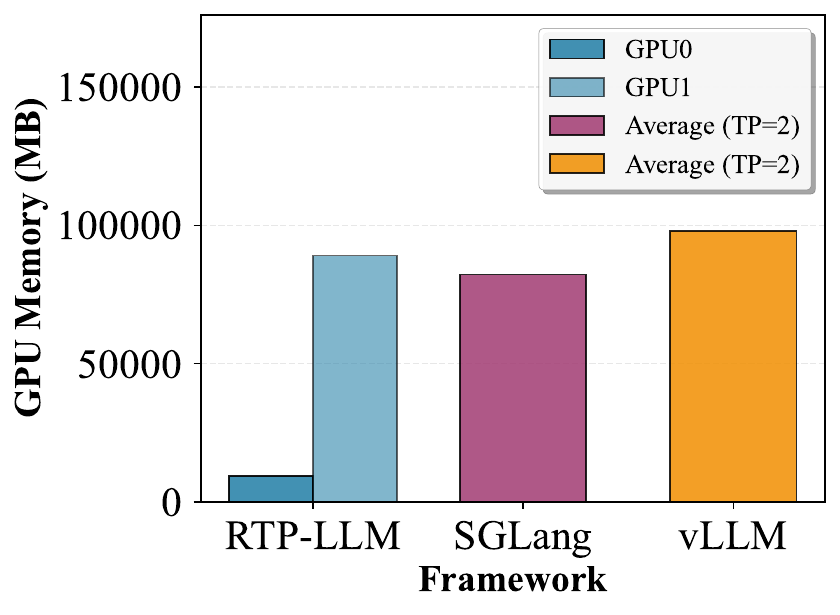}}{(d) GPU Memory}
    \caption{Performance and GPU memory utilization comparison for Qwen/Qwen2.5-VL-7B-Instruct on GQA dataset across different frameworks. All frameworks are deployed with TP=2. RTP-LLM shows GPU0 and GPU1 memory usage separately, while SGLang and vLLM show average memory usage across both GPUs.}
    \label{fig:vit_decoupling_performance}
    \vspace{-3mm}
\end{figure*}

\subsection{Quantized Inference Performance}
{Throughput and latency are measured under controlled request patterns; precision (PPL) is evaluated on the public WikiText-2 \cite{mikasenghaasWikitext2} dataset. We evaluate RTP-LLM's quantization performance for Qwen3-32B model across different quantization configurations (AWQ(FP8), FP8 KV Cache, Baseline) \cite{qwen3technicalreport}, comparing against SGLang and vLLM. {Here, Baseline denotes Qwen3-32B without KV cache quantization and AWQ quantization}. All runs use single GPU deployment (TP = 1, DP = 1) and consistent parameters: max\_batch\_size = 64, top-$p$ = 1, top-$k$ = 1, temperature = 0.0. The performance metrics include Batch Latency, TTFT, Tokens/s, and PPL on WikiText-2. We vary max\_new\_tokens over 500, 800, 1000, and 1200.}

Figure~\ref{fig:quantization_batch_latency} presents the batch latency and precision loss comparison across different quantization configurations. RTP-LLM demonstrates superior batch latency performance in Baseline configuration (Figure~\ref{fig:quantization_batch_latency}(c)), achieving the lowest batch latency across all max\_new\_tokens settings. For FP8 KV Cache configuration (Figure~\ref{fig:quantization_batch_latency}(b)), RTP-LLM achieves the lowest batch latency for longer sequences (800, 1000, 1200 max\_new\_tokens), demonstrating better scalability compared to baseline frameworks. RTP-LLM achieves significant batch latency reduction of 35\%-40\% in FP8 KV Cache configuration compared to AWQ(FP8) configuration (Figure~\ref{fig:quantization_batch_latency}(a)), while maintaining competitive performance in AWQ(FP8) configuration.

Figure~\ref{fig:quantization_ttft} presents TTFT and tokens/s results across different configurations. RTP-LLM achieves the lowest TTFT across all the configurations, demonstrating 1.9x-3.0x reduction compared to SGLang and vLLM. %Figure~\ref{fig:quantization_tokens_per_sec} presents the tokens/s comparison across all the quantization configurations.
In addition, RTP-LLM demonstrates high performance across all the quantization configurations, achieving the highest tokens/s in AWQ(FP8) and Baseline configurations. In FP8 KV Cache configuration, RTP-LLM maintains competitive performance with minimal differences compared to baseline frameworks, while consistently achieving superior batch latency for longer sequences and TTFT. The superior performance is due to RTP-LLM's superior engineering implementation, including optimized quantization kernel  and efficient memory access patterns.

% {These performance advantages stem from RTP-LLM's superior engineering implementation, including optimized quantization kernel implementations and efficient memory access patterns.}

{Figure~\ref{fig:quantization_batch_latency}(d) shows precision loss (PPL) on a sampled subset of WikiText-2. All configurations yield PPL in 7.59--8.09 (baseline 7.59--7.67). RTP-LLM attains the best precision in Baseline and comparable precision in AWQ(FP8) and FP8 KV Cache (within 0.01 PPL of baseline).}

% RTP-LLM demonstrates slightly lower GPU memory usage compared to SGLang and vLLM across all quantization configurations, though the differences are minimal and not significant enough to warrant detailed presentation.

\vspace{-3mm}
\subsection{EPD Disaggregation Performance}

The evaluation is motivated by production vision-language workloads such as {\textbf{structured product description generation}} (e.g., for second-hand marketplace listings): each request contains {\textbf{multiple product images}} and an {\textbf{image-retrieval-based reference product list}}, requiring multi-image understanding and coherent long-form text generation. To enable reproducible framework comparison, we use the \textbf{public GQA benchmark} \cite{hudson2019gqanewdatasetrealworld}, which similarly demands both visual and textual understanding and is representative of such multi-image, visual-question-answering-style tasks. We evaluate RTP-LLM's Vision Transformer (ViT) decoupling (EPD) performance for the Qwen/Qwen2.5-VL-7B-Instruct model \cite{qwen2.5-VL, Qwen2VL, Qwen-VL} on GQA, comparing against SGLang and vLLM. All runs use Tensor Parallelism (TP=2) deployment across two GPUs and consistent parameters: Max\_batch\_size = 64, max\_new\_tokens = 500, top-$p$ = 1, top-$k$ = 1, temperature = 0.0, ensuring fair comparison across different frameworks.

Figure~\ref{fig:vit_decoupling_performance} presents the performance and GPU memory utilization comparison across three frameworks. RTP-LLM demonstrates superior performance across all metrics. In terms of throughput (Figure~\ref{fig:vit_decoupling_performance}(c)), RTP-LLM achieves 6288.48 tokens/s, which is 1.86x faster than SGLang (3374.24 tokens/s) and 2.52x faster than vLLM (2492.69 tokens/s). In terms of the latency metric, RTP-LLM achieves the lowest TTFT at 1737.48 ms (Figure~\ref{fig:vit_decoupling_performance}(b)), representing 2.36x and 2.12x reduction compared to SGLang (4103.1 ms) and vLLM (3688.3 ms), respectively. RTP-LLM also achieves the lowest time cost at 3823.24 ms (Figure~\ref{fig:vit_decoupling_performance}(a)), reducing latency by 20.1\% and 21.5\% compared to SGLang (4784 ms) and vLLM (4874 ms), respectively. This superior performance stems from RTP-LLM's decoupled architecture, which enables overlap between ViT processing and language model execution---particularly beneficial for production workloads with {multi-image input and image-retrieval reference context}, where vision encoding and text generation can be pipelined efficiently.

Figure~\ref{fig:vit_decoupling_performance}(d) presents the GPU memory utilization comparison. RTP-LLM's EPD Disaggregation architecture enables efficient memory distribution across GPUs: GPU0 uses only 9{,}279.81 MB while GPU1 uses 89{,}087.81 MB, demonstrating significant memory savings on the first GPU. In contrast, SGLang and vLLM distribute memory more evenly across both GPUs under TP=2 deployment, with average memory usage of 82{,}210.34 MB (average of 82{,}442.12 MB and 81{,}978.56 MB across two GPUs) and 97{,}871 MB (average across two GPUs), respectively. This asymmetric memory distribution in RTP-LLM is achieved through the decoupled design that separates vision encoding (processed on GPU0 with minimal memory footprint) from language generation (processed on GPU1), resulting in better resource utilization and reduced computational overhead, as evidenced by the superior throughput and latency metrics.

\vspace{-2mm}
\section{Conclusions}
\label{sec:conclusion}
We present RTP-LLM, a high-performance and production-ready inference engine designed to meet the rigorous demands of industrial-scale LLM deployment within Alibaba Group. RTP-LLM optimizes production-scale model loading to enable rapid elastic scaling, achieves superior memory efficiency via adaptive quantization and KV cache management, implements robust distributed processing using the Prefill-Decode Disaggregation architecture, and crucially, deploys effective production traffic scheduling strategies to maximize GPU utilization and guarantee Service Level Objectives (SLOs) under complex, high-concurrency workloads. 
Experimental results demonstrate that RTP-LLM delivers substantial advantages in throughput and latency compared to state-of-the-art open-source baselines, using both public benchmarks and {real production workloads}; the system successfully powers mission-critical applications across various Alibaba business units. RTP-LLM is released as open-source software, and the design choices, configurations.
%, and evaluation methodology reported in this paper are applicable to other large-scale LLM deployment environments beyond our own infrastructure. 
%The system serves as a foundational infrastructure for high-efficiency, reliable LLM deployment.
%\section{Future Works}
%\label{sec:future_works}
%Looking ahead, we are committed to maintaining RTP-LLM’s position at the forefront of LLM serving technology by integrating several advanced architectural and model innovations into our production environment. 
In future, 
%we plan to further optimize the compute pipeline efficiency by launching the \textbf{AFD Disaggregation} architecture (Attention and FFN Disaggregation) %\cite{bhatia2025helixparallelismrethinkingsharding, singh2025efficientlyservinglargemultimodal} 
%to achieve more fine-grained computational decoupling. 
%Furthermore, our engine is being prepared for the immediate, full-scale deployment of next-generation ultra-large models, notably the \textbf{Qwen3-Next} %\cite{Qwen3NextBlog} 
%series. 
we plan to further explore acceleration techniques such as the \textbf{DSA} (DeepSeek Sparse Attention) mechanism. %\cite{deepseekai2024deepseekv32}, 
%aiming to leverage its sparsity benefits to achieve the most extreme computational and memory bandwidth gains. %\textbf{Additionally, we will continue refining our industrial-grade request scheduler to achieve dynamic, latency-aware load balancing across heterogeneous resources.} %This ensures RTP-LLM continues to drive the industrial adoption of cutting-edge AI while maintaining stability.

\bibliographystyle{ACM-Reference-Format}
\bibliography{myref}

\end{document}